\documentclass{aa}  
\usepackage[dvipsnames]{xcolor} 
\usepackage{graphicx,txfonts}
\usepackage[breaklinks=true]{hyperref} 
\hypersetup{colorlinks=true,citecolor=Blue}
\usepackage{natbib}
\bibpunct{(}{)}{;}{a}{}{,}

\usepackage[english]{babel}
\usepackage{blindtext}
\usepackage{pifont}
\usepackage[normalem]{ulem}
\usepackage{comment}
\usepackage{placeins}
\usepackage{amsmath}
\usepackage{caption} 
\usepackage{makecell}
\usepackage{subcaption}
\usepackage{multirow}
\usepackage{appendix}
\usepackage{gensymb}

\begin{document} 

   \title{Survival of Protoplanetary Disks in Upper Scorpius from Population Synthesis Models with External Photoevaporation}

   \author{Jingyi Ping \inst{1}\thanks{\email{jingyi.ping@student.kuleuven.be}}
        \and
        Rossella Anania \inst{2}
        \and 
        Paola Pinilla \inst{3}
        \and
        Miguel Vioque \inst{4}
          }

   \institute{Institute of Astronomy, KU Leuven, Celestijnenlaan 200D, 3001 Leuven, Belgium
   \and
    Dipartimento di Fisica, Università degli Studi di Milano, Via Celoria 16, I-20133 Milano, Italy 
    \and
    Mullard Space Science Laboratory, University College London, Holmbury St Mary, Dorking, Surrey RH5 6NT, UK
    \and
    European Southern Observatory, Karl-Schwarzschild-Str. 2, 85748 Garching bei München, Germany}
   \date{}
 
  \abstract
  {We present population synthesis models of viscous protoplanetary disks subject to mild external far-ultraviolet (FUV) radiation fields ($F_{\mathrm{UV}}=1$–100\,G$_0$). Our simulations focus on gas disk evolution, exploring stellar masses drawn from an Initial Mass Function and a range of initial disk conditions. We quantify the fraction of surviving disks across 10\,Myr of evolution, track the evolution of gas disk mass and size, and compare our results with observations of protoplanetary disks in the Upper Scorpius region, including the ten targets studied by the AGE-PRO ALMA Large Program. We find that models combining viscous evolution with external photoevaporation yield disk lifetimes of 3–7\,Myr, consistent with observed dispersal timescales, particularly for $10^{-4}\leq\alpha\leq10^{-2}$. Low-mass stars ($0.1$\,M$_\odot$) are more susceptible to disk dispersal due to their weaker gravitational binding, with their fraction among all surviving disks dropping from 76\% at birth to 51\% by 10\,Myr. The majority of the long-lived disks are those with low viscosity $\alpha<10^{-3.5}$ and initial characteristic radius $R_c<125$\,AU; while the initial disk-to-star mass ratio does not play an important role. The median gas disk mass and radius of the surviving disks exhibit a sharp decline in the first 0.2\,Myr of evolution, followed by a slight increase that reflects survivorship bias. We also explore correlations between gas disk mass and size vs. stellar mass and FUV strength. Our findings highlight the critical role of external photoevaporation in shaping disk populations even at moderate levels of FUV radiation fields.}

\titlerunning{Population Synthesis Models with External Photoevaporation} 
\maketitle

\section{Introduction}
Protoplanetary disks are the birth sites of planets. Their evolution and dispersal strongly affect the final properties of planets. There are two physical processes that can transport angular momentum in protoplanetary disks: viscous transport mediated by turbulence \citep{Shakura1973_alpha,Lynden1974} and angular momentum removal driven by magneto-hydrodynamical (MHD) winds \citep{Blandford1982, Bai2013}. Both of these scenarios leave clear imprints in the disk properties. For example, purely viscous models lead to an increase in gas disk size over time, while MHD winds are predicted not to increase the disk radius with the disk evolution \citep{yang2021, Trapman2022}. Distinguishing the dominant mechanisms between these two from observations remains one of the biggest challenges in the planet formation community \citep{Manara2023,Zhang2025, Tabone2025}.

There are additional physical processes that can shape disk evolution, such as photoevaporation, where high energy radiation from the central star  \citep[internal;][]{Clarke2001,Ercolano2008,Picogna2019,Sellek2024} or nearby massive stars \citep[external; for a review see][]{Winter2022,Allen2025} heat the protoplanetary disks and trigger a thermal wind when the thermal velocity of particles is higher than the escape velocity. Although photoevaporation does not transport angular momentum, it can remove the disk material, modify the disk size and accelerate disk dispersal \citep[e.g.,][]{Clarke2007,Coleman2022}. External photoevaporation originates from the ultraviolet (UV) radiation emitted by massive stars, mainly nearby O and B stars, and the disk fraction decreases with the number of O/B stars in clusters \citep{Guarcello2023}.

Mild UV radiation in clusters can also affect the evolution of protoplanetary disks, as recently studied for protoplanetary disks in Upper Scorpius (hereafter Upper Sco) by \cite{anania2025}. Upper Sco is located at an average distance of $\sim$ 142 pc from the Sun \citep{Fang2023} and is one of the nearest star-forming regions. It has an age of $\sim$ 3–14\,Myr \citep[e.g.,][]{Ratzenbock2023} and exemplifies the late stages (Class II) of disk evolution \citep{Pecaut2012, Armstrong2025}. Upper Sco hosts $\sim 50$ OB stars \citep{Luhman2020} and 284 stars with full, evolved, transitional and debris disks are identified in \citet{carpenter2025}. Upper Sco serves as a perfect laboratory to investigate the fate and survival of disks that are exposed to the moderate far-ultraviolet (FUV) flux (with values of $F_{\mathrm{UV}}\sim1$–100\,G$_0$) from the OBA-type stars in the region.  Former population synthesis models including viscous evolution and internal photoevaporation cannot fully explain the accretion rate and the gas disk mass of the ALMA Survey of Gas Evolution of PROtoplanetary disks (AGE-PRO) sample of both Lupus (1–3\,Myr) \citep{Deng2025} and Upper Sco (2–6\,Myr) disks \citep{Agurto-Gangas2025}, favoring MHD wind-driven models \citep{Tabone2025}. The results of the AGE-PRO collaboration suggest that the gas mass distribution is similar between Lupus and Upper Sco, implying that the surviving disks in Upper Sco have gas disk masses similar to their younger counterparts \citep{Trapman2025, Zhang2025}. However, the above models cannot fully recover gas disk sizes in Lupus and Upper Sco. \citet{anania2025} modeled the same Upper Sco disks with viscous evolution models that include external photoevaporation, explaining better the gas disk sizes and masses than when including only viscous evolution and neglecting external photoevaporation in the models. This demonstrates that, due to the role of the stellar formation environment, Upper Sco disks cannot be regarded as old counterparts of Lupus disks \citep{Zagaria2023}. Moreover, \citet{Coleman2024} demonstrated that even weak external photoevaporation (similar $F_{\mathrm{UV}}$ values to those experienced by disks in Upper Sco, i.e., $\sim$ 10\,G$_0$) can dampen the effectiveness of disk radius as a diagnostic for distinguishing between viscous and wind-driven disks. Together, these results highlight the importance of constraining disk evolution using population synthesis models that incorporate external photoevaporation and enable direct comparison with observations.

In this work, we aim to investigate the role of moderate external photoevaporation ($F_{\mathrm{UV}}=1$–100\,G$_0$) in shaping the disk fraction and the evolution of gas disk masses and sizes of a population of protoplanetary disks. We explore whether a gas-only population synthesis model of viscous disks experiencing external photoevaporation can reproduce the distribution of gas disk sizes observed in the Upper Sco region. Moreover, we aim to investigate whether the correlations from our population synthesis models align with disks in Upper Sco observed in the AGE-PRO sample, and provide potential observational correlations that can be used as a test of our models with future observations of a larger sample in Upper Sco.

The paper is structured as follows. In Sect. \ref{Sec:target_sample}, we explain the observational sample used in this work to compare with the results from models. In Sect. \ref{Sec:gas_evol_model}, we describe the gas evolution model, simulation setup and the method used to compare with observations. Section \ref{Sec:results} presents the main results obtained from the simulations. We focus on the disk fraction, gas disk mass, size, and empirical correlations for gas disk mass and size. Our results and additional processes affecting disk evolution are discussed in Sect. \ref{Sec:discussion}. Section \ref{Sec:conclusions} summarizes the main remarks.

\section{Target sample}\label{Sec:target_sample}
The targets considered for comparing the results of our population synthesis to observations are the 202 disks in Upper Sco reported by \citet{carpenter2025}. The disks were initially identified through near- and mid-infrared excess emission detected with WISE and Spitzer \citep{Esplin2018,Luhman2020,Luhman2022}, yielding a total of 284 candidates, of which 202 were subsequently identified as Upper Sco members.

\begin{figure}[!htbp]
  \centering
  \includegraphics[width=0.47\textwidth]{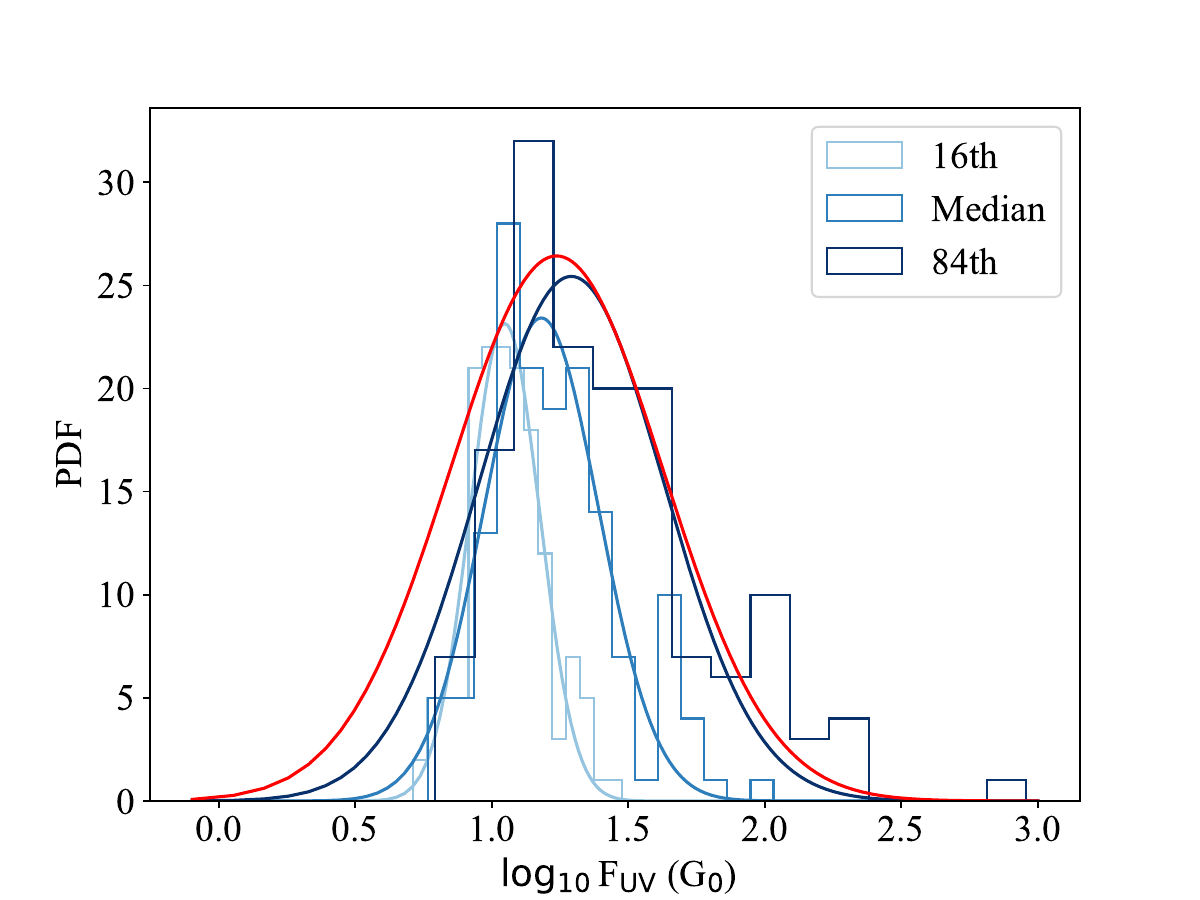} 
  \caption{PDF of external FUV flux $F_{\mathrm{UV}}$, fitted with log normal distribution. Light, medium and dark blue lines represent 16th, median and 84th value of the FUV flux in the Upper Sco star forming region, respectively. The red line enclosing the blue lines is the distribution from which we draw FUV flux $F_{\mathrm{UV}}$ in our models.}
  \label{Fig:fuv_distribution}
\end{figure}

The FUV fluxes used in our population synthesis models are sampled from the distribution of values related to the observed sample. The FUV fluxes $F_{\mathrm{UV}}$ at the position of the Upper Sco targets are evaluated by considering the contribution of the neighboring OBA-type stars, and accounting for the uncertainty in the 3D separation from massive stars, using the method detailed in \citet{Anania2025fuv}. Figure \ref{Fig:fuv_distribution} demonstrates the probability density function (PDF) of the observed FUV flux with the 16th, median, 84th values and their corresponding log-normal fits (blue). The red represents the log-normal distribution that encloses these observed values, from which the FUV fluxes in our models are drawn. Approximately 72\% of the disks are subject to mild level of FUV flux $F_{\mathrm{UV}}$ between 3 and 30 G$_0$. 

\begin{figure}[!htbp]
  \centering
  \includegraphics[width=0.47\textwidth]{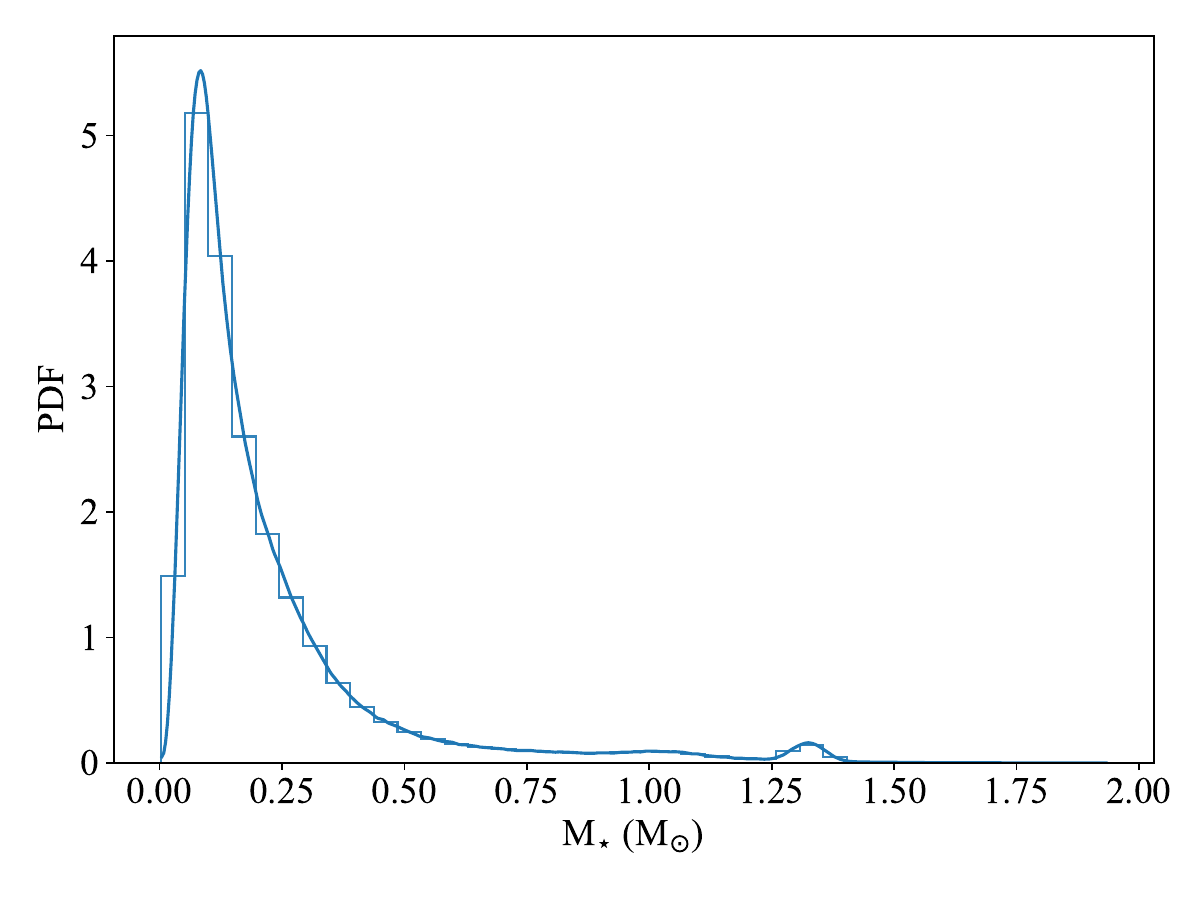} 
  \caption{PDF of stellar mass $M_{\star}$ for the 169 disk-bearing stars in Upper Sco by \citet{carpenter2025}. The stellar masses are presented in \cite{Pinilla2025}. }
  \label{Fig:stellar_mass_distribution}
\end{figure}

In addition, to compare the simulation results with the observed sample, we show the PDF of stellar mass of those disk-bearing stars in Figure \ref{Fig:stellar_mass_distribution}. Stellar masses are available for 169 of the 202 disks, adopted from \cite{Pinilla2025}, who derived them from stellar luminosities and effective temperatures by comparison with pre–main-sequence evolutionary tracks from \citet{Baraffe2015}. The remaining 33 targets lack luminosity measurements and thus have no stellar mass estimates but do not show significant bias in spectral type, and thus in mass. Among the 169 disks with mass estimates, about 65\% of the stars have $M_\star < 0.2$\,M$_\odot$, and no disk-bearing star more massive than 3\,M$_\odot$ has been identified. Based on the spectral energy distribution and near-infrared color–color diagram, 97 disks are classified as “full”, 14 as “transitional”, 26 as “evolved” and 23 as “debris”, “evolved transitional” or class III by \citet{Luhman2020,Luhman2022}.

As a comparison for the gas disk size distribution resulting from our model, we used the definition of radius as in \citet{carpenter2025}, which is the radius enclosing 95\% of the $^{12}$CO J = 3-2 flux emission in the image plane. Due to the resolution of the observations, the radius has a lower limit of $0.25 \arcsec$ in \citet{carpenter2025} ($\sim$ 36\,AU at the Upper Sco distance of 142\,pc). Therefore, 148 disks with radii at this limit are considered unresolved, leaving 54 disks with gas disk size measurements.

\section{Gas evolution models}\label{Sec:gas_evol_model}
We perform  population synthesis models of gas-only simulations using \texttt{DustPy} \citep{Stammler2022} along with an integrated module  that incorporates external photoevaporation \citep{Garate2024, anania2025} by interpolating the FRIEDv2 grid \citep{Haworth2023} to low values of $F_{\mathrm{UV}}$ (down to $1\,\mathrm{G_0}$) and including the mass loss rate by external photoevaporation into the models as explained in \cite{Garate2024}.

Considering a geometrically thin and axisymmetric disk evolving under viscous accretion and external photoevaporation, the gas surface density $\Sigma_{\mathrm{g}}\left(R,t\right)$ can be obtained by the 1D standard diffusion equation \citep{Lynden1974,Pringle1981}:
\begin{equation}
  \dfrac{\partial }{\partial t} \Sigma_{\mathrm{g}} = \dfrac{3}{R}\dfrac{\partial}{\partial R} \left(R^{1/2}\dfrac{\partial}{\partial R}\left( \nu\Sigma_{\mathrm{g}}R^{1/2}\right)\right) - \dot{\Sigma}_{\mathrm{ext}},\
  \label{eq:lynden}
\end{equation}
where $\nu =\alpha c_s H_g$ is the kinetic viscosity, with $H_g$ being the gas scale height, $c_s$ the sound speed, and $\alpha$  a dimensionless viscosity parameter \citep{Shakura1973_alpha}. In Eq.~\ref{eq:lynden},  $\dot{\Sigma}_{\mathrm{ext}}$ is the mass loss rate due to external photoevaporative winds. This mass loss rate is derived via the bilinear interpolation of the FRIEDv2 grid \citep{Haworth2023}. 
The ratio of Polycyclic aromatic hydrocarbon (PAH) to the dust abundance is assumed to be 1, corresponding to ISM-like PAH for ISM-like dust. The PAH-to-dust ratio is important, as PAH is the dominant heating mechanism in the photodissociation region (PDR) \citep{Facchini2016}. In the absence of observational constraints, we adopt a fiducial value of 1. In the external photoevaporation models, the truncation radius $R_{\mathrm{t}}$ is defined as the radius with the maximum mass loss rate, marking the transition between the optically thick and optically thin regions of the externally photoevaporative wind. The total mass loss rate outside the truncation radius ($R > R_{\mathrm{t}}$) $\dot{M}_{\mathrm{tot}}$ is then predicted for a disk with a given surface density profile around a star with certain mass subject to FUV radiation field. Following \cite{Sellek2020}, the total mass loss rate is distributed across a grid cell ($i$), weighted by the total mass in the outer region:
\begin{equation}
\dot{M}_{\mathrm{ext, \, i}} =\dot{M}_{\mathrm{tot}}\dfrac{M_{\mathrm{i}}}{M(R > R_{\mathrm{t}})}.
\end{equation}

In addition, we consider a passive flared disk, heated by the central star under an angle $\psi$, the temperature profile is
\begin{equation}
  T(R) = \left(\dfrac{\psi L_{\star}}{8 \pi R^2 \sigma_{\mathrm{SB}}}\right)^{1/4},
\end{equation}
where $\sigma_{\mathrm{SB}}$ is the Stefan-Boltzmann constant, and $L_{\star}$ is the stellar
luminosity, which is obtained for different stellar masses based on the evolutionary tracks by \citet{Dotter2008} for stars at the age of 5\,Myr. In Appendix \ref{app:variable_l}, we show a comparison with the scenario in which the stellar luminosity $L_{\star}$ evolves according to the evolutionary track by \citet{Dotter2008}, showing that the results are analogous to the scenario of fixed luminosity. A constant flaring irradiation angle $\psi =0.05$ is adopted, leading to the temperature profile of $T(R) \propto R^{-1/2}$.

\subsection{Initial conditions}
The initial surface density profile is set based on the \citet{Lynden1974} self-similar
solution, assuming a linear dependence of the viscosity $\nu \propto R^{\gamma}$ with $\gamma =1$,
\begin{equation}
\Sigma_g(R)=\Sigma_0 \left(\dfrac{R}{R_c}\right)^{-1}\exp \left(-\dfrac{R}{R_c}\right),
\end{equation}
where $R_\mathrm{c}$ is the characteristic radius, the scale radius of the exponential term, and $\Sigma_0$ is set by the total disk mass.

The disk is modeled with a radial grid that consists of 500 cells uniformly spaced in $R^{1/2}$, covering a range from 2.5 to 1000\,AU. 

\subsection{Population synthesis models}
A total of 5000 simulations are performed, covering well the entire relevant parameter space. The primary parameters considered include: stellar mass $M_{\star}$, disk-to-stellar mass ratio ($\epsilon$= $M_{\mathrm{disk}}/M_{\star}$), disk characteristic radius $R_\mathrm{c}$, viscosity parameter $\alpha$ and external FUV flux $F_{\mathrm{UV}}$. Table \ref{Tab:disk_initial_parameters} shows the ranges of the parameters considered in the population synthesis models.

The stellar mass $M_{\star}$ is selected based on the initial mass function (IMF) of \citet{Maschberger2013}, which is built upon the IMF of \citet{Kroupa2001,Kroupa2002} and \citet{Chabrier2003}. Stellar masses are randomly sampled over the 0.1–3\,M$_\odot$ range covered by the FRIEDv2 grid. Similarly, the FUV field strengths, restricted to the discrete values available in the FRIEDv2 grid, are sampled according to relative probabilities based on the observed log-normal distribution (the red line in Figure \ref{Fig:fuv_distribution}). The other initial parameters, viscosity parameter $\alpha$, disk-to-stellar mass ratio $\epsilon$ and disk characteristic radius $R_\mathrm{c}$ are randomly drawn from a log uniform distribution. 

\begin{table}[!htbp]
  \centering
  \caption{Initial Disk Parameters}
  \label{Tab:disk_initial_parameters}
  \renewcommand{\arraystretch}{1.2} 
  \setlength{\tabcolsep}{3.5pt}    
  \footnotesize                    
  \begin{tabular}{lccc} 
    \hline\hline
    Parameter & Description & Range & PDF \\
    \hline
    $\alpha$ & Viscosity & $10^{-5}$ – $10^{-2}$ & Log uniform \\
    $M_\star$ (M$_{\odot}$) & Stellar mass & \makecell[l]{0.1, 0.3, 0.6,\\1.0, 1.5, 3.0} & IMF \\
    $\epsilon = M_{\mathrm{disk}} / M_{\star}$ & Disk-to-stellar mass ratio & $10^{-3}$ – 0.5 & Log uniform \\
    $R_\mathrm{c}$ (AU) & Characteristic radius & $10$ – $200$ & Log uniform \\
    $F_{\mathrm{UV}}$ ($\mathrm{G_0}$) & External FUV flux & 1, 10, 100 & Log normal \\
    \hline
  \end{tabular}
\end{table}

Each disk is evolved for 10\,Myr. We compare results in which the disk is considered to be dispersed once having lost 90, 95 and 99\% of the initial mass, or when the mass loss rate drops below $10^{-14}$ M$_{\odot}$ yr$^{-1}$. In our cases, all disks are dispersed by the mass-based criteria before the mass loss rate reaches the threshold.

\subsection{Comparison with observations}\label{Sec:compar_with_obs}
We assume that gas disk radius is the position at which the gas column density encloses 90\% of $^{12}$CO line flux. The gas column density $N_{\mathrm{gas}}(R_{^{12}\mathrm{CO},\,90\%})$ can be written as a function of the gas disk mass $M_{\mathrm{gas}}$ and the stellar luminosity $L_{\star}$ as  in \citet{Trapman2023}:
\begin{equation}
  N_{\mathrm{gas}}\left(R_{^{12}\mathrm{CO}, \, 90\%} \right) \approx 10^{21.27-0.53 \log L_{\star}}\left(\dfrac{M_{\mathrm{gas}}}{\mathrm{M_{\odot}}} \right)^{0.3-0.08 \log L_{\star}},
\end{equation}
which is obtained from fitting the results of DALI thermodynamical models and relies on the initial carbon abundance and the assumed disk temperature profile. 

In our models, $R_{^{12}\mathrm{CO}, \, 90\%}$ is calculated as the radius corresponding to the critical value of the disk surface density at each time step,
\begin{equation}
\Sigma(R_{^{12}\mathrm{CO}, \, 90\%} , \,t)=\Sigma_{\mathrm{crit}}(t) = N_{\mathrm{gas}}(R_{^{12}\mathrm{CO}, \, 90\%} \, ,\, t) \mu_{\mathrm{gas}},
\end{equation}
where $\mu_{\mathrm{gas}}$ is the mean molecular weight of the gas in the disk, which is set to be $\sim 2.3 \, \mu_{\mathrm{H}} = 1.15 \, \mu_{\mathrm{H_2}}$.

\section{Results and comparison with observations}\label{Sec:results}
\subsection{Disk fraction and surviving disks}\label{subsec:disk_fraction_and_surviving}
\begin{figure}[!htbp]
  \centering
  \includegraphics[width=0.47\textwidth]{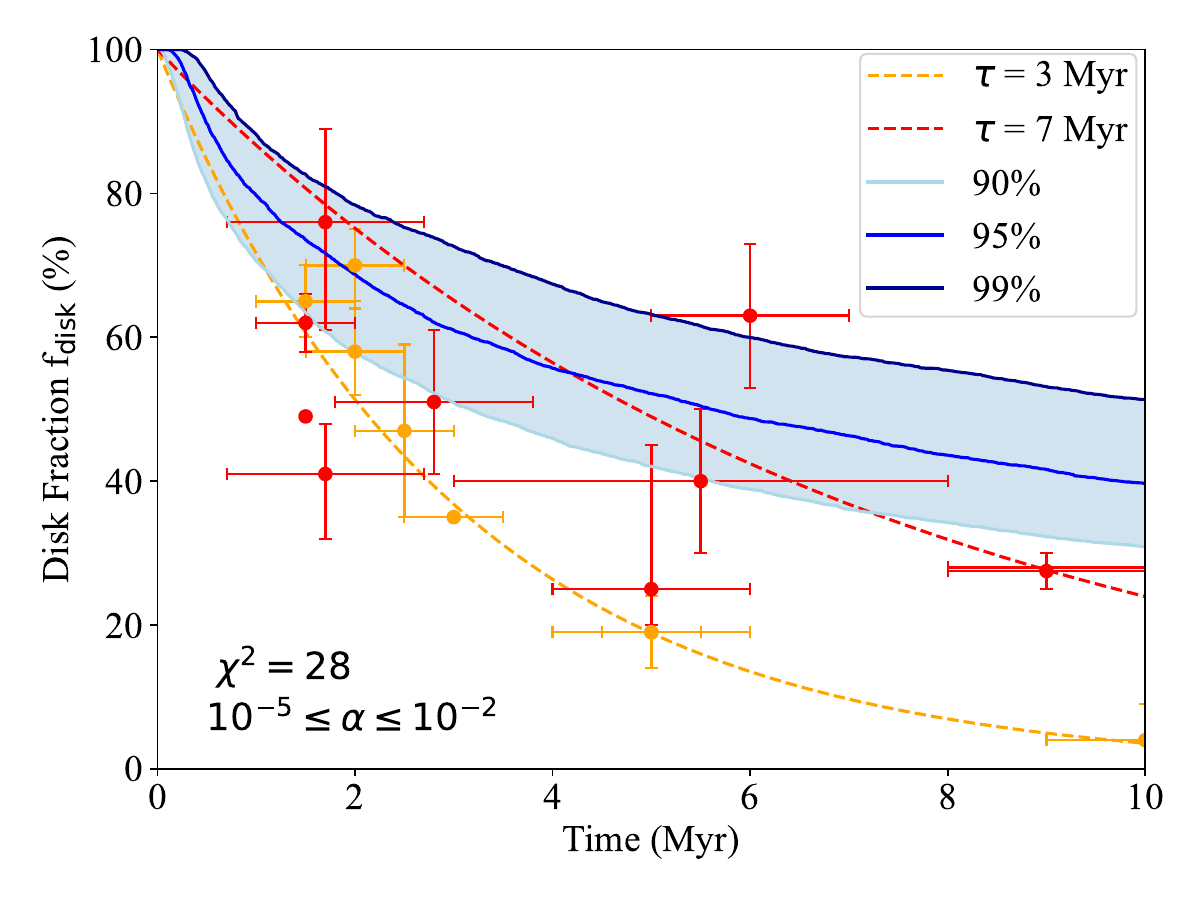}
  \vspace{1em}
  \includegraphics[width=0.47\textwidth]{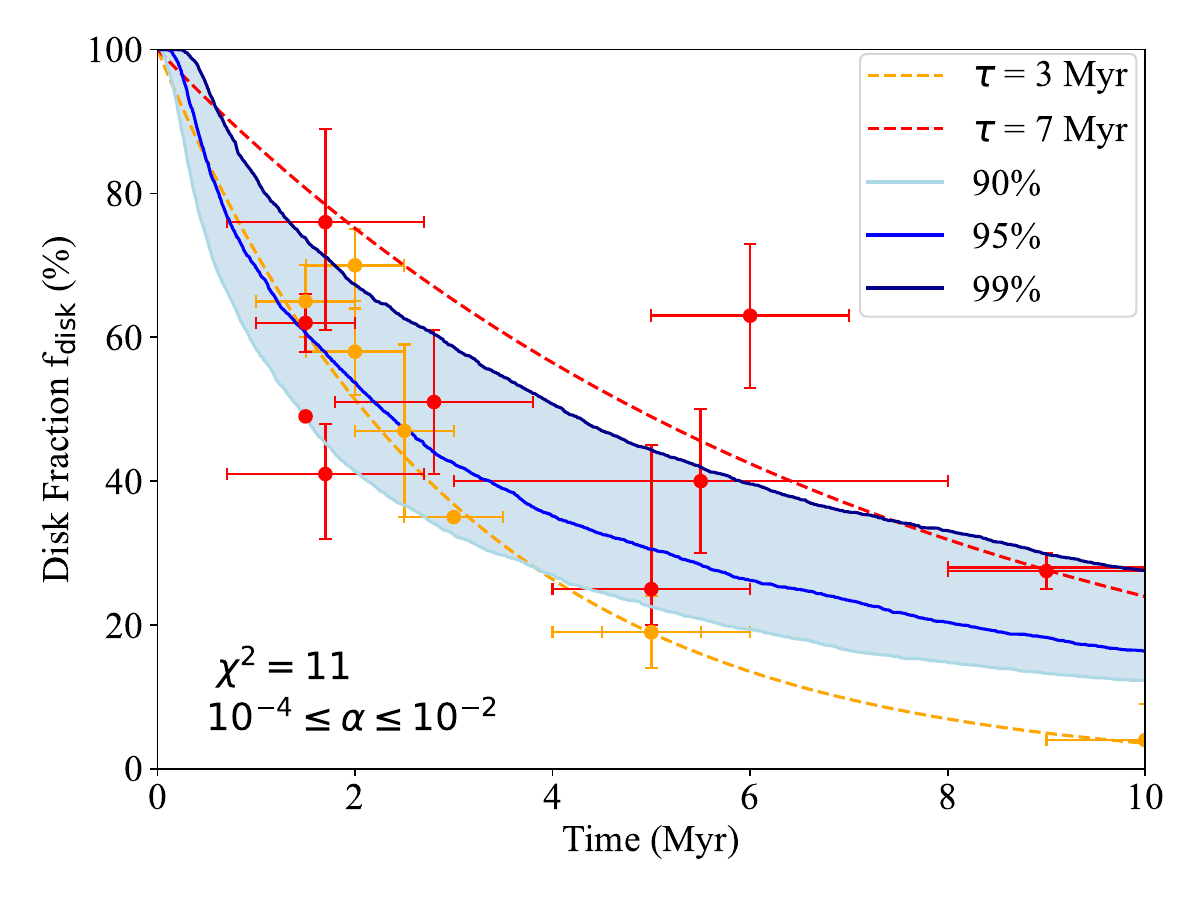}
  \caption{The decline in disk fraction $f_{\mathrm{disk}}$ with disk age. Blue lines are the disk fraction considering disk dispersal when losing 90\% (light), 95\% (medium) and 99\% (dark) of the initial mass. The red points correspond to the observed disk fractions of nearby clusters within 200 pc, and the dashed line is the fitted exponential decay, $f(t)= \exp(-t/\tau)$, yielding $\tau =7$\,Myr from  \cite{Pfalzner2022}. The yellow points are from \citet{Fedele2010}, giving $\tau =3$\,Myr. The top and bottom panel correspond to disks with viscosity ranges of $10^{-5} \leq \alpha \leq 10^{-2}$ and $10^{-4} \leq \alpha \leq 10^{-2}$, respectively.  The $\chi^2$ values are shown in the bottom left of each figure. }
  \label{Fig:disk_fraction}
\end{figure}

Figure \ref{Fig:disk_fraction} shows the decline in disk fraction with disk age from our population synthesis models (blue), in comparison with different disk lifetimes. Different blue lines represent gas disk dispersal when losing different percentages of the initial gas disk mass. The red and orange lines show the observed disk fractions and the corresponding derived disk lifetime of $\tau = 3$\,Myr \citep{Fedele2010} and $\tau = 7$\,Myr \citep{Pfalzner2022}, assuming an exponential decay $f(t)= \exp(-t/\tau)$. The disk lifetime is constrained by the accretion timescale, reflecting gas dispersal, and by the infrared excess timescale, tracing dust dissipation. These timescales depend on stellar mass, wavelength, evolutionary models, and the environment \citep{Mamajek2009, Ribas2015, Fang2025}. The accretion timescale is typically shorter than the infrared excess timescale, with characteristic values $\sim$ 3\,Myr \citep{Fedele2010,Delfini2025}. \citet{Mamajek2009} estimated the infrared excess timescale $\sim$ 2.5\,Myr. But this can extend to $\tau \sim 7$\,Myr when only considering nearby star-forming regions ($d < 200$\,pc) that lack massive O-type stars \citep{Pfalzner2022}.

In our simulations, disk fraction $f_{\mathrm{disk}}$ decreases more slowly than the observed values with viscosity $10^{-5} \leq \alpha \leq 10^{-2}$ (top panel of Figure \ref{Fig:disk_fraction}). At 5\,Myr, the disk fraction considering disk dispersal when losing 95\% of the initial mass is $\sim 6\%$ above the empirical relation with $\tau =7$\,Myr. Later at 10\,Myr, the disk fraction assuming disk dispersal when losing 90\% of the initial mass is $\sim 29\%$  larger than the empirical relation with $\tau = 7$\,Myr. 

However, when restricting to disks with $10^{-4} \leq \alpha \leq 10^{-2}$, the resulting time evolution of disk fraction is more consistent with observations, as shown in the bottom panel of Figure \ref{Fig:disk_fraction}. This is because higher viscosity increases the accretion rate onto the central star, resulting in shorter disk lifetimes. Also, higher viscosity drives faster viscous spreading, making disks more extended. Therefore, disks with higher viscosity are more vulnerable to external photoevaporation. Other different ranges of viscosity are also considered in appendix \ref{app:disk_frac}. Among them, the disk fraction favors viscosity $10^{-4} \leq \alpha \leq 10^{-2} $ to match with observations, when assuming disk viscous evolution under external photoevaporation.

\begin{figure}[!htbp]
  \centering
  \includegraphics[width=0.47\textwidth]{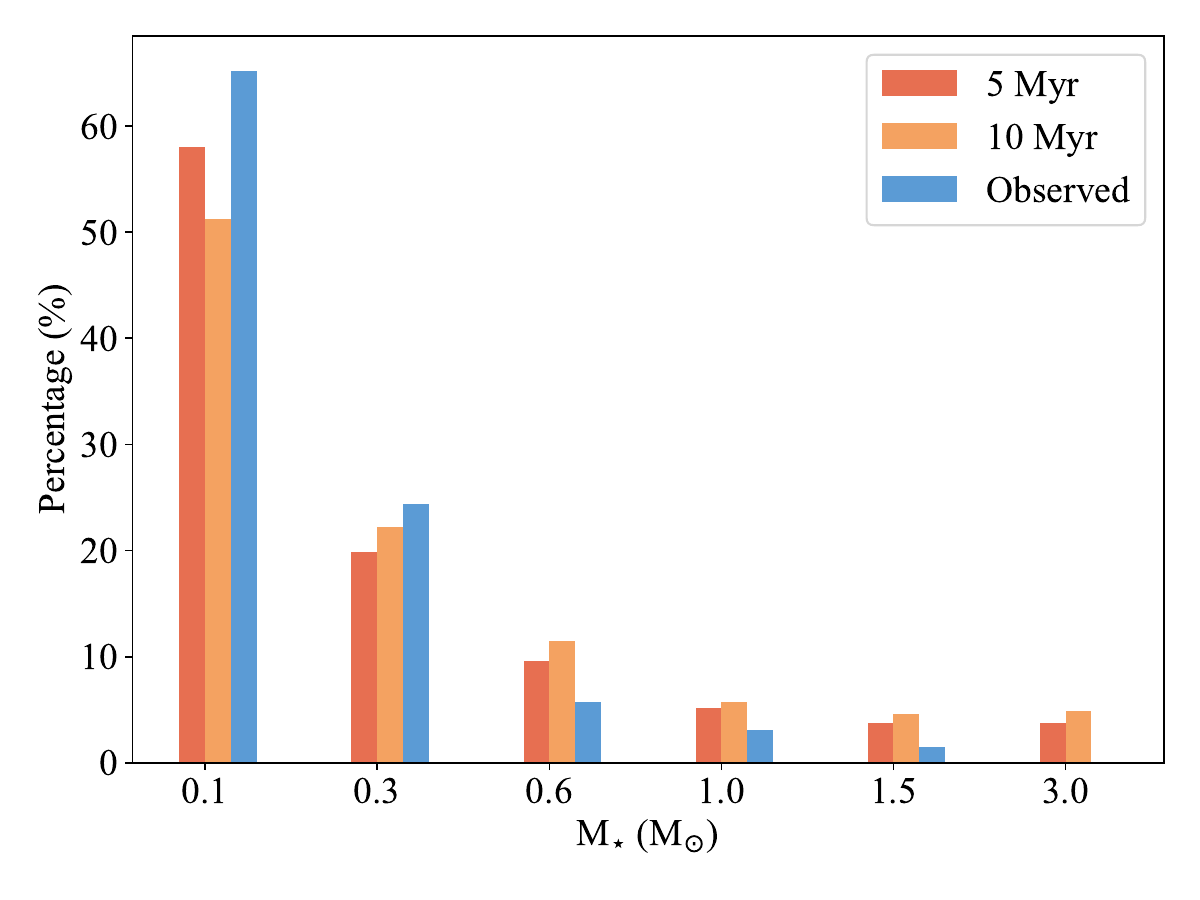} 
  \caption{The percentage of surviving disks in each stellar mass bin from our simulations at 5 (red) and 10 (orange)\,Myr of disk evolution, compared to the 169 samples of \citet{carpenter2025} (blue). The stellar masses are presented in \cite{Pinilla2025}. }
  \label{Fig:survive_mstar}
\end{figure}

We check the distribution of the stellar mass of the surviving disks, when disk dispersal is defined as losing 90\% of the initial gas disk mass (the same criteria unless noted otherwise). Figure \ref{Fig:survive_mstar} shows the percentage of surviving disks in each stellar mass bin at different times of disk evolution, compared to observations. Material in disks around less massive stars is less gravitationally bound, hence easier to be entrained by photoevaporative winds. As a result, the percentage of surviving disks around 0.1 M$_{\odot}$ stars among all surviving disks decreases over time, while that of disks around more massive stars gradually increases. The fraction of surviving disks around $0.1$\,M$_{\odot}$ decreases from $\sim$ 76\% at birth to around 51\% at 10\,Myr. A similar trend is observed for stellar mass, with more disk-bearing stars found at lower masses. However, observation suggests that the percentage of disks around stars with 0.1 M$_{\odot}$ is $\sim$ 27\% higher than in our simulations. No star with $3$\,M$_{\odot}$ is present in the observed sample. Indeed, for stars more massive than $1.5$\,M$_\odot$, photoevaporation by the central star is expected to disperse the disk by 10\,Myr of evolution \citep{Kunitomo2021}. Consequently, internal photoevaporation, which is neglected in our models, will change the stellar mass distribution of the surviving disks.

\begin{figure}[!htbp]
  \centering
  \includegraphics[width=0.47\textwidth]{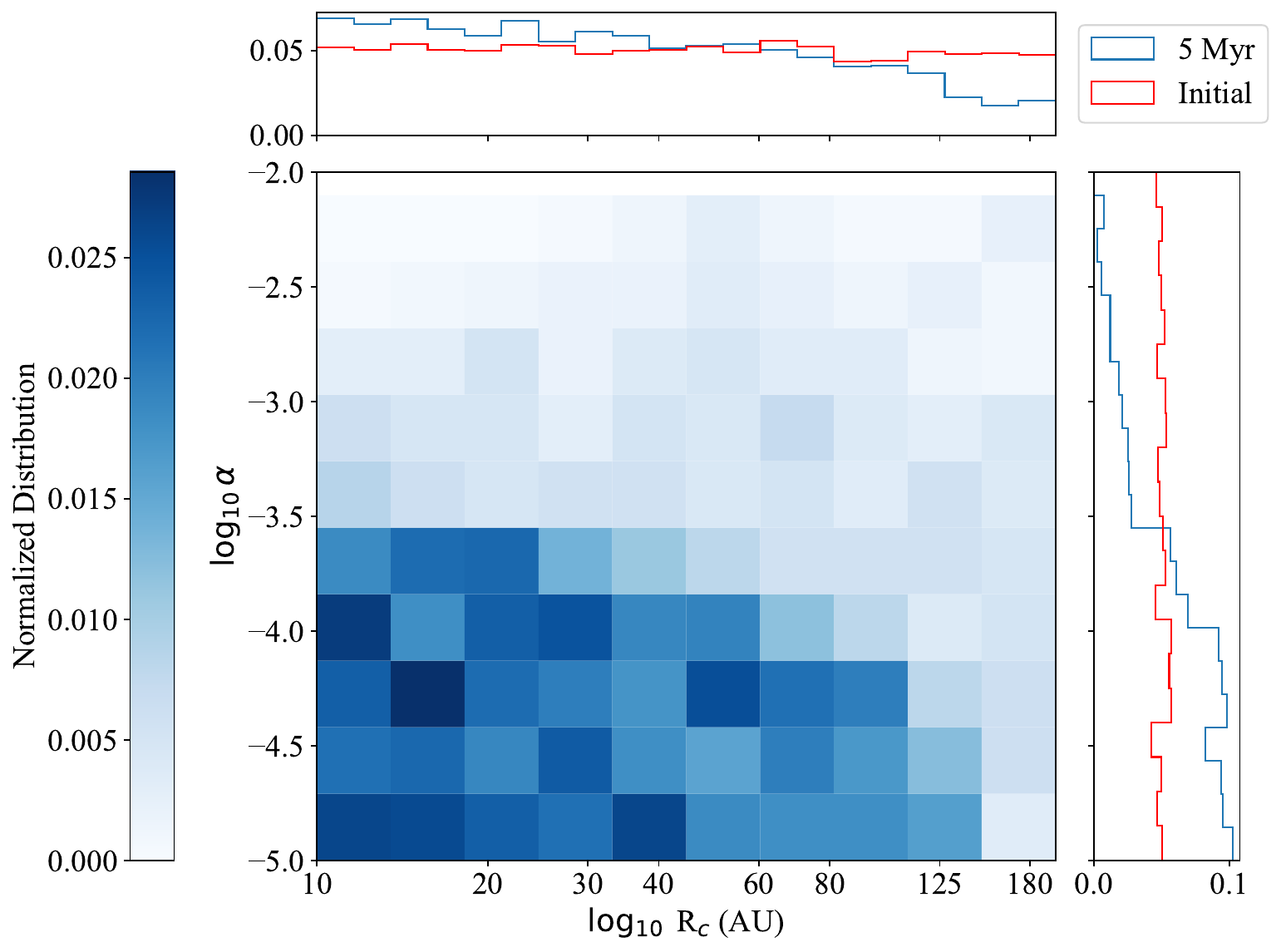} 
  \caption{The distribution of disks with certain characteristic radius $R_{\mathrm{c}}$ and viscosity parameter $\alpha$ that survive after 5\,Myr of evolution, assuming that gas disk is  dispersed when losing 90\% of the initial mass. The red and blue represent the distributions at birth and 5\,Myr, respectively. }
  \label{Fig:survive_alpha_rc}
\end{figure}

To understand the properties of the surviving disks in our models, Figure \ref{Fig:survive_alpha_rc} shows the distribution of disks that survive after 5\,Myr with a given value of $R_{\mathrm{c}}$ and $\alpha$. The surviving disks exhibit a preference for low viscosity and small characteristic radius $R_{\mathrm{c}}$, where the preference for low viscosity is stronger than that for small characteristic radius. The surviving disks at 5\,Myr mainly exhibit viscosity $\alpha < 10^{-3.5}$, whereas no significant preference is observed for characteristic radius $R_{\mathrm{c}} < 125$\,AU. Lower viscosity reduces the efficiency of disk spreading, and therefore, a combination of lower viscosity $\alpha$ and smaller $R_{\mathrm{c}}$ results in more compact disks that are more gravitationally bound and less vulnerable to mass loss by external photoevaporation. 

\subsection{Gas disk mass evolution}\label{subsec:gas_disk_mass_evol}
\begin{figure}[htbp]
  \centering
  \includegraphics[width=0.47\textwidth]{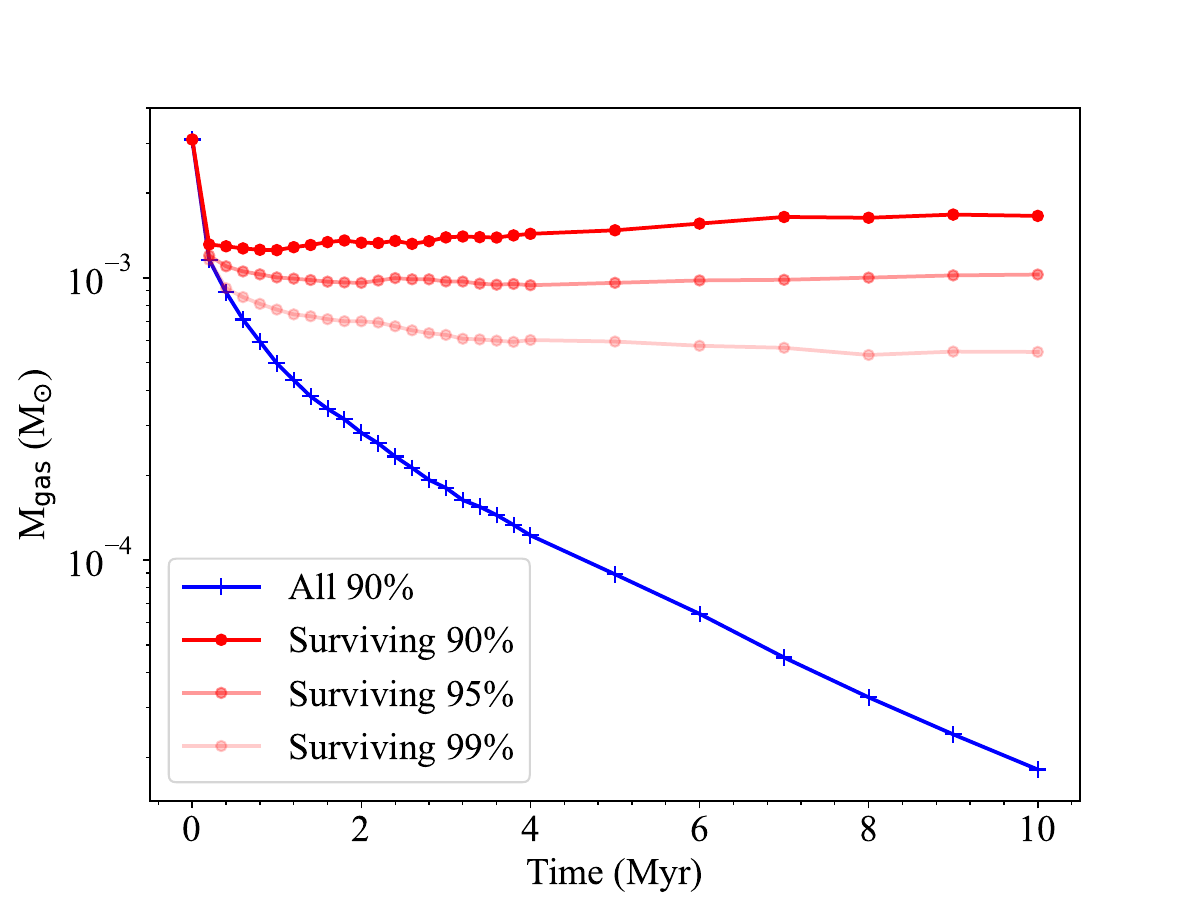} 
  \caption{Time evolution of the median gas disk mass $M_{\mathrm{gas}}$. The blue line shows the evolution of the entire sample, adopting 90\% gas disk mass loss as the dispersal criterion. The red lines refer to surviving disks only, with dark, medium, and light red corresponding to dispersal defined by 90\%, 95\%, and 99\% of the gas disk mass loss, respectively.}
  \label{Fig:gas_disk_mass}
\end{figure}

Figure \ref{Fig:gas_disk_mass} shows the time evolution of the median gas disk mass of the surviving disks, which decreases sharply within the first 0.2\,Myr and increases slightly thereafter, as the low mass disks disperse. If we consider disk dispersal when losing 90\% of the initial gas disk mass, the median gas disk mass decreases by $\sim 58\%$ at the first 0.2\,Myr. Later at 10\,Myr, the median gas disk mass is $\sim 26\%$ larger than that at 0.2\,Myr. A similar trend is observed as well if we consider disk dispersal when losing 95 or 99\% of the initial mass. In contrast, the median value of the entire disk population continues to decline over time. This highlights the importance of survivorship bias among the observed disk population.

\begin{figure}[htbp]
  \centering
  \includegraphics[width=0.47\textwidth]{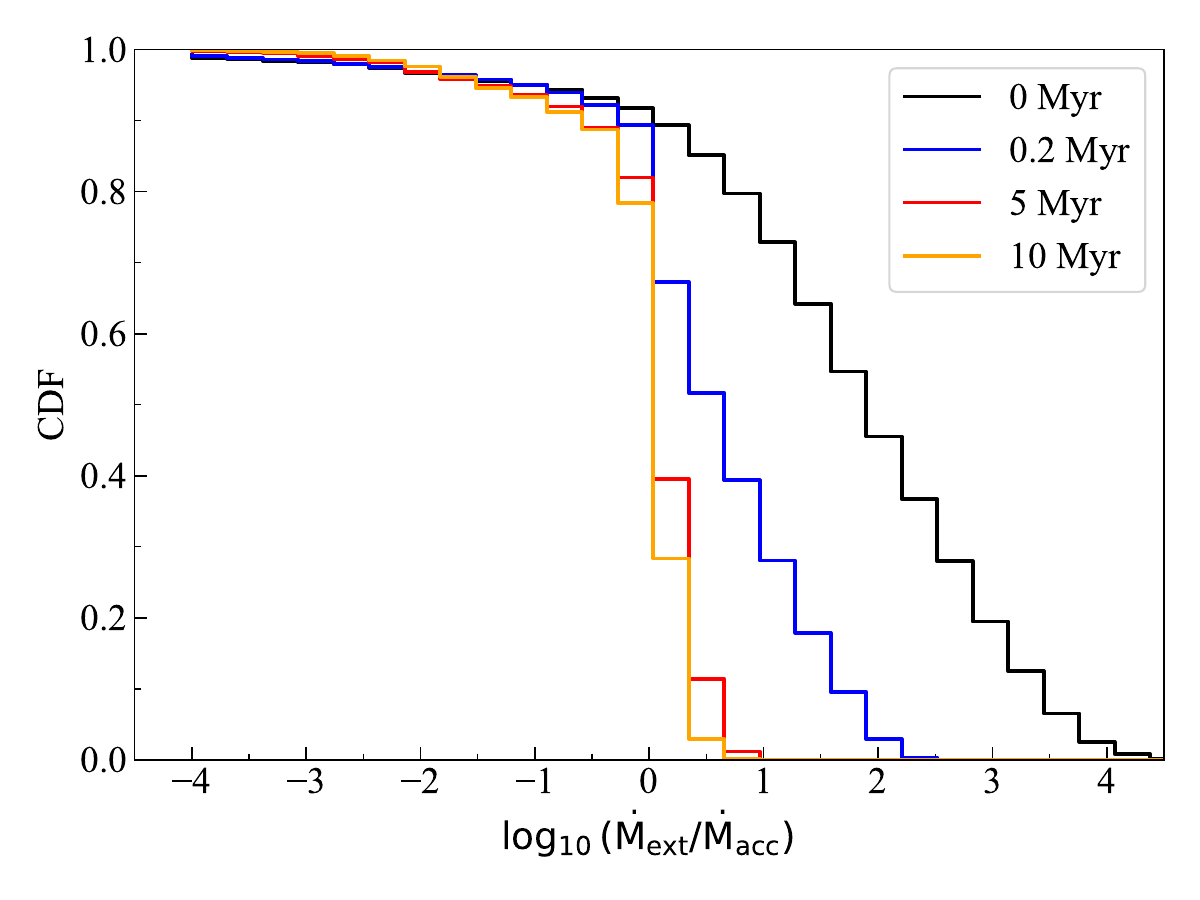} 
  \caption{Cumulative Distribution Function (CDF) of $\dot{M}_{\mathrm{ext}}/\dot{M}_{\mathrm{acc}}$. The black, blue, red and orange correspond to the values at 0, 0.2, 5 and 10\,Myr of disk evolution, respectively.}
  \label{Fig:mloss_macc}
\end{figure}

The gas mass evolution of the disk is mainly influenced by the mass accreted onto the central star and the mass removed by external photoevaporative winds. At the beginning of the simulation, $\sim 80\%$ of disks show a mass loss rate at least one order of magnitude larger than the mass accretion rate, as illustrated in Figure \ref{Fig:mloss_macc}. Our results suggest that for the whole population, mild FUV flux ($F_{\mathrm{UV}} =1$–100\,G$_0$) can cause rapid decline of the median disk mass in a relatively short timescale of 0.2\,Myr. As the disk evolves, the accretion rate onto the star gradually reaches the equilibrium with the rate of materials lost in winds \citep{Clarke2007,Sellek2020}. After 0.2\,Myr, $\sim 38\%$ of the surviving disks exhibit mass loss rate at least one order of magnitude larger than the mass accretion rate. Later at 5\,Myr, this fraction drops to only $\sim 0.8\%$.

Moreover, the surviving disks do not show a significant preference for higher initial disk-to-star mass ratios $\epsilon$. The slightly higher initial masses observed among surviving disks arise from the stronger dispersal of disks around low-mass stars. Although disks with lower $\epsilon$ are easier to be dispersed, this effect remains marginal. 

\subsection{Gas disk size}\label{subsec:gas_disk_size}
\begin{figure}[htbp]
  \centering
  \includegraphics[width=0.47\textwidth]{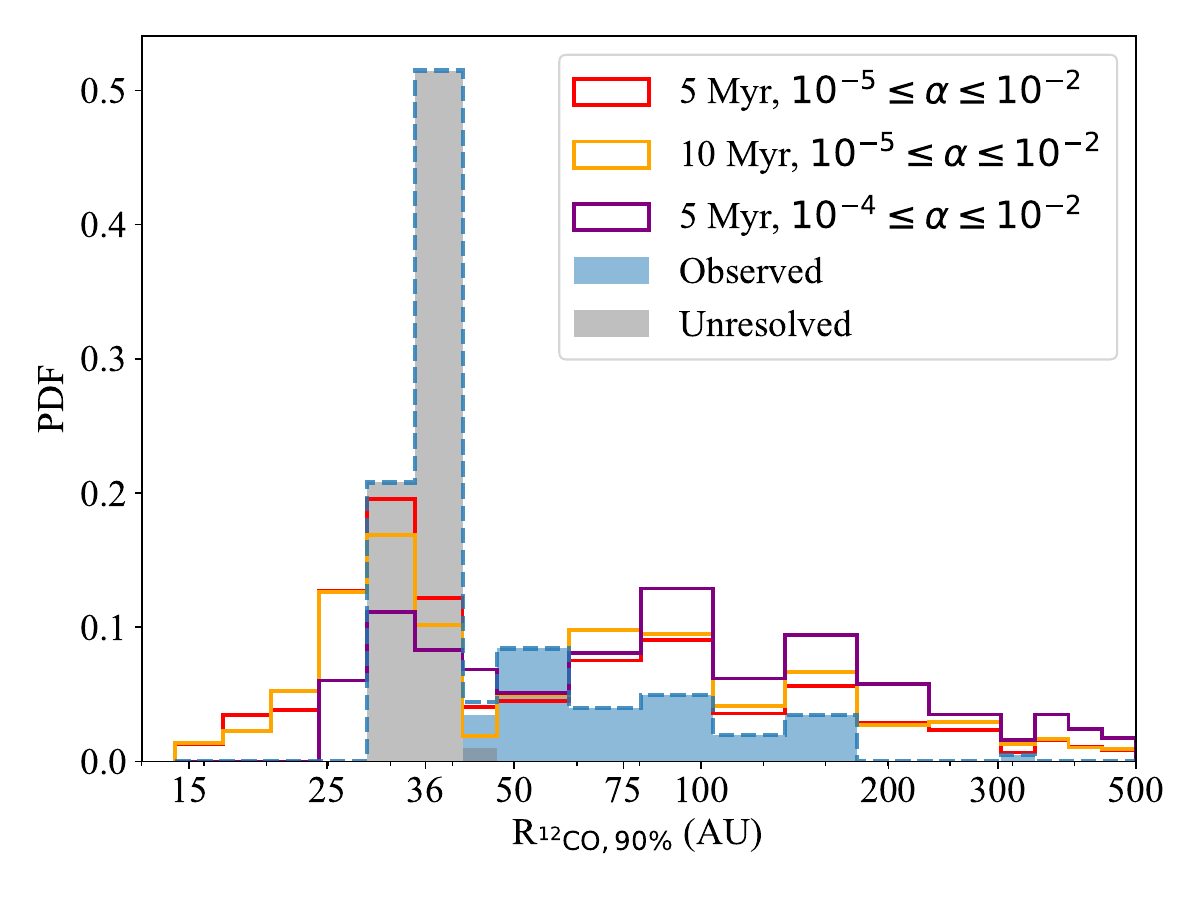} 
  \caption{The PDF of the gas disk radius $R_{^{12}\mathrm{CO}, \, 90\%}$ for surviving disks at 5 (red) and 10 (orange)\,Myr, compared with observation by \citet{carpenter2025} (blue). The gray marks the unresolved disks with upper limit radius $0.25\arcsec$, and the dashed blue line includes all the disks. The purple line selects models with viscosity $10^{-4} \leq \alpha \leq 10^{-2}$.}
  \label{Fig:disk_Rco}
\end{figure}

Figure \ref{Fig:disk_Rco} shows the PDF of gas disk radius $R_{^{12}\mathrm{CO}, \, 90\%}$ compared with observations by \citet{carpenter2025} (similar to \citet{Zallio2025}). Both viscosity $10^{-4} \leq \alpha \leq 10^{-2}$ and $10^{-5} \leq \alpha \leq 10^{-2}$ can produce extended disks with radii $> 150$\,AU. However, surviving disks with higher viscosity remain large under external photoevaporation, due to efficient viscous spreading. In \citet{carpenter2025}, 148 disks ($\sim 73\%$ of the total sample) remain unresolved, with an upper limit radius of $0.25 \arcsec$ ($\sim$ 36\,AU at 142\,pc) imposed by the beam size. \citet{Zallio2025} further showed that the median gas disk radius is expected to be $\lesssim$ 22 \,AU. The higher viscosity range $10^{-4} \leq \alpha \leq 10^{-2}$ fails to reproduce this large population of compact disks. Under our initial conditions, viscosity $10^{-4} \leq \alpha \leq 10^{-2}$ only produces $\sim 18\%$ of disks with radius < 36 AU, whereas lower viscosity $10^{-5} \leq \alpha \leq 10^{-2}$ produces $\sim 44\%$. In contrast to the disk fraction, which favors viscosity of $10^{-4} \leq \alpha \leq 10^{-2}$, the gas disk radius distribution prefers lower viscosity to produce more compact disks, which constitute a high fraction of the observed population. Resolving the size of the small disks that remain unresolved by current observations will provide new insights about the disk conditions that may reproduce the observed disk properties.

\begin{figure}[!htbp]
  \centering
  \includegraphics[width=0.47\textwidth]{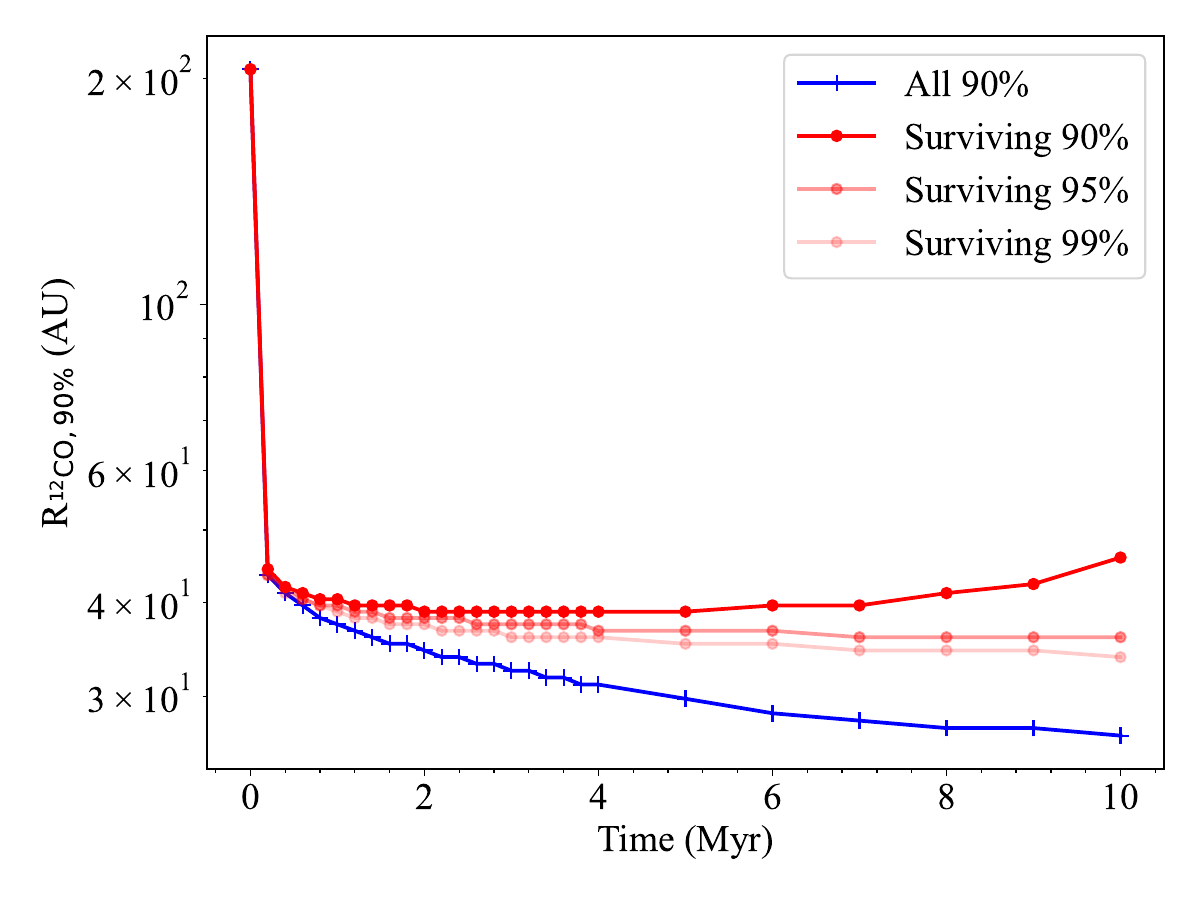} 
  \caption{Time evolution of the median gas disk radius $R_{^{12}\mathrm{CO},\,90\%}$. Same layout as Figure \ref{Fig:gas_disk_mass}. }
  \label{Fig:disk_Rco_time}
\end{figure}

In Figure \ref{Fig:disk_Rco}, the PDF of the surviving disks only shows a slight difference between 5 and 10\,Myr. Figure \ref{Fig:disk_Rco_time} illustrates the time evolution of the median gas disk radius. The evolution of the median gas disk radius follows a similar trend to the gas disk mass, declining to $\sim 21\%$ of the initial value within the first 0.2\,Myr, remaining nearly constant until 5\,Myr, and increasing slightly thereafter. At 10\,Myr, the median radius is approximately 18\% larger than at 5\,Myr. Survivorship bias is evident for the gas disk radius as well.

As discussed above, disk mass loss due to external photoevaporation is already significant across the entire population from the early stages of evolution. External photoevaporative winds truncate the disk rapidly within the first 0.2\,Myr. Larger disks are easier to be truncated, while initially compact disks viscously spread until the mass loss rate becomes significant, then shrink. 

Similar trends can also be seen from disk radius $R_{90\%}$ that encloses 90\% of the gas disk mass, the median value of radius decreases significantly during the first 0.2\,Myr, remains almost constant afterwards, then increases after 9\,Myr, but now only slight differences between the overall population and the surviving disks. The disks are truncated by the external photoevaporative winds to similar sizes. The gas disk radius $R_{^{12}\mathrm{CO},\,90\%}$ is comparable to or larger than the radius $R_{90\%}$, and can reach up to five times $R_{90\%}$.

\subsection{Empirical correlations}\label{subsec:expected_corr}
\begin{figure*}[t!]
  \centering
  \includegraphics[width=\linewidth]{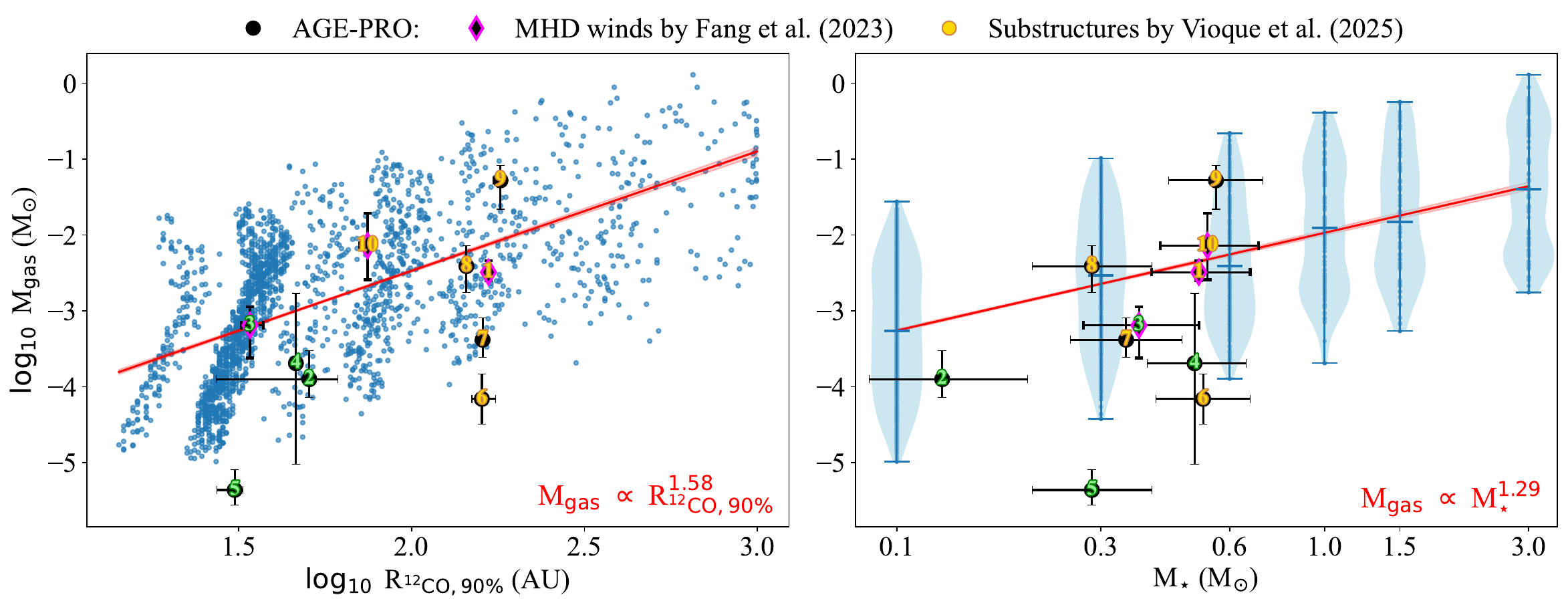}
  \caption{Empirical correlations of the gas disk mass with  gas disk radius (left) and stellar mass (right) at 5\,Myr. The blue marks values derived from simulations and the black is the values of 10 targets in AGE-PRO. The red line is fitted using a MCMC method, and the 1$\sigma$ credible interval is marked in light red. Disks with substructures identified in AGE-PRO \citep{Vioque2025} are shown in yellow, and those with MHD winds traced by [O I] $\lambda$ 6300 line \citep{Fang2023}, are highlighted in bright purple.}
  \label{Fig:corr_mgas}
\end{figure*}

\begin{figure*}[t!]
  \centering
  \includegraphics[width=\linewidth]{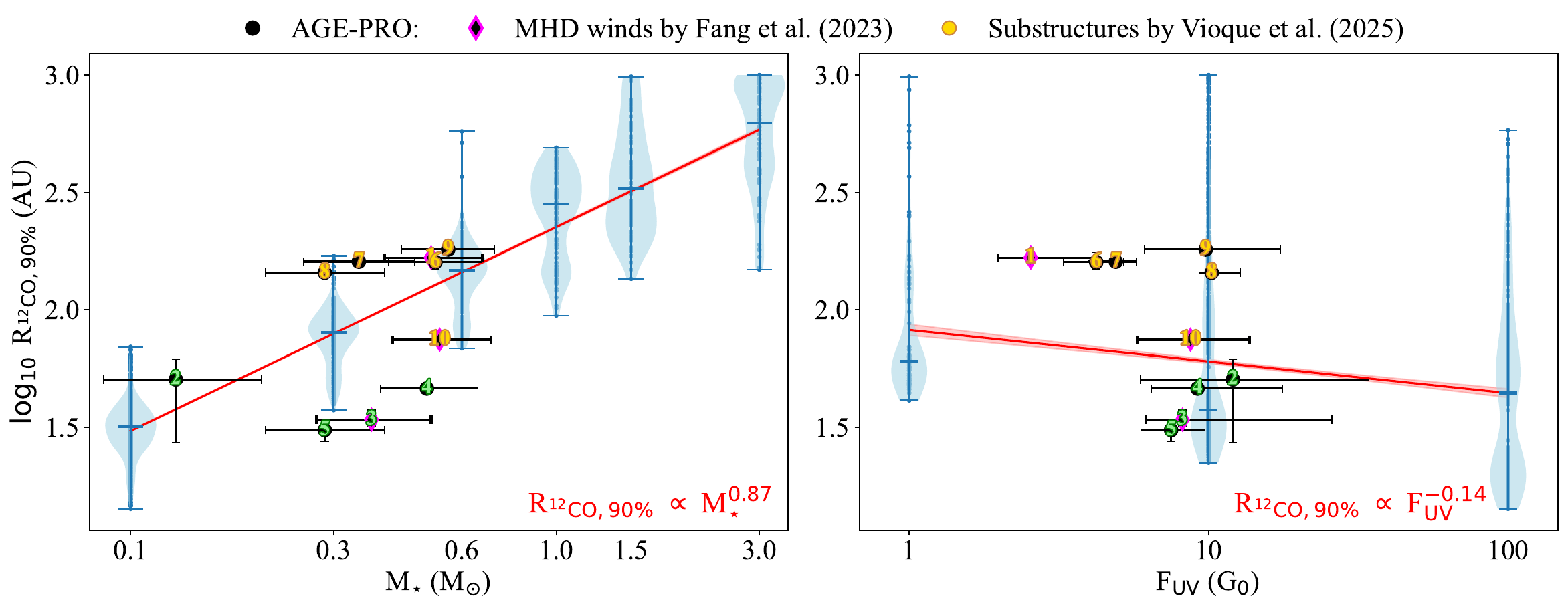}
  \caption{Empirical correlations of the gas disk radius with stellar mass (left) and FUV flux (right) at 5\,Myr, following the same format as Figure \ref{Fig:corr_mgas}.}
  \label{Fig:corr_rco}
\end{figure*}

We search for potential correlations between the properties of surviving disks and stellar properties.
We apply a Markov Chain Monte Carlo (MCMC) method using \texttt{linmix} \citep{Kelly2007} to estimate the Pearson correlation coefficient r between different parameters. All the surviving disks from our simulations are included in the analysis. We then fit the parameters with a power-law relation. The median value and 1$\sigma$ confidence interval (CI) of the slope $\alpha$ and the intercept $\beta$ are calculated. The main results are summarized in Figure \ref{Fig:corr_mgas} and \ref{Fig:corr_rco}, while a full list of the correlation coefficients can be found in Table \ref{Tab:correlation}.  
We find that the gas disk mass $M_{\mathrm{gas}}$ presents a moderate positive correlation ($r \sim 0.6$) with gas disk size $R_{^{12}\mathrm{CO}, \, 90\%}$, a tentative positive correlation ($r \sim 0.5$) with stellar mass $M_{\star}$, and a negligible correlation ($r \sim 0.04$) with FUV field $F_{\mathrm{UV}}$. For the gas disk radius, a very strong correlation ($r \sim 0.9$) with stellar mass and a negligible correlation ($r \sim -0.2$) with FUV strength are found. 

Figures \ref{Fig:corr_mgas} and \ref{Fig:corr_rco} illustrate these correlations for the surviving disks at 5\,Myr of evolution. We do not apply any uncertainty to our simulated data, and therefore, the resulting 1$\sigma$ credible intervals from the MCMC fitting are correspondingly narrow. We obtain $M_{\mathrm{gas}} \propto R_{^{12}\mathrm{CO},\,90\%}^{1.58}$ and $M_{\mathrm{gas}} \propto M_{\star}^{1.29}$, $R_{^{12}\mathrm{CO},\,90\%} \propto M_{\star}^{0.87}$ and $R_{^{12}\mathrm{CO},\,90\%} \propto F_{\mathrm{UV}}^{-0.14}$. Additionally, the power-law changes slightly over time. For example, the power-law index decreases by $\sim$ 14\% from 1\,Myr to 10\,Myr for correlations between $M_{\mathrm{gas}}$ and $R_{^{12}\mathrm{CO},\,90\%}$. Also, there is large scatter in the gas disk mass (as illustrated in Figure \ref{Fig:corr_mgas}) due to the combination of different initial parameters. This scatter makes it challenging to identify such correlations from a limited observational sample. 

Similar to gas disk mass, the scatter in gas disk radius is huge, as shown in Figure \ref{Fig:corr_rco}. The strongest correlation is found between stellar mass and gas disk radius. The observed decrease in gas disk radius with increasing FUV flux is consistent with the expectation that stronger FUV radiation can truncate disks more effectively. 

\begin{figure}[!htbp]
  \centering
  \includegraphics[width=0.47\textwidth]{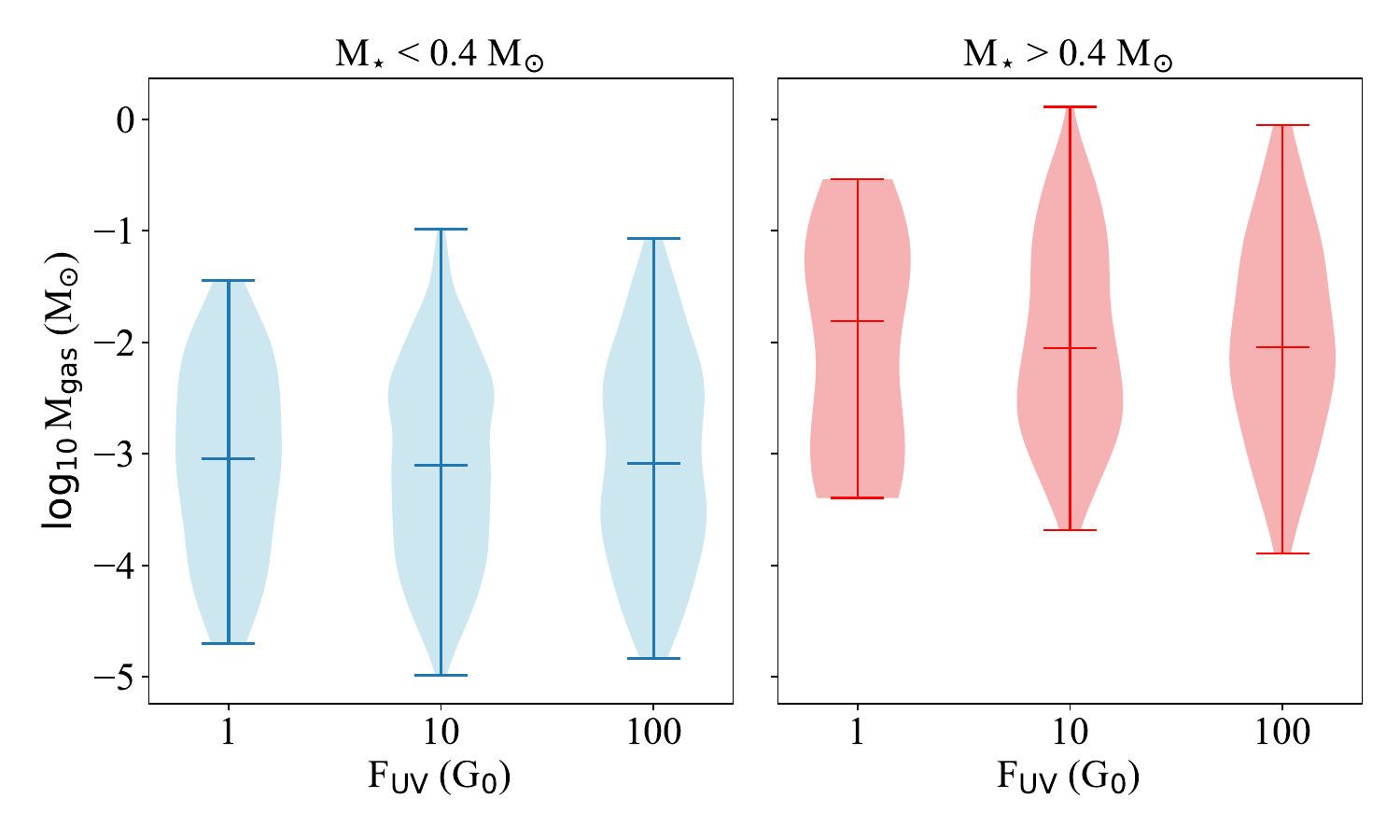} 
  \caption{The distribution of gas disk mass over FUV strength of surviving disks at 5\,Myr. The blue and red represent disks around low-mass stars with $M_{\star} < 0.4$\,M$_{\odot}$ and high-mass stars with $M_{\star} > 0.4$\,M$_{\odot}$.}
  \label{Fig:corr_mgas_fuv}
\end{figure}

However, the dependence of gas disk radius on FUV flux does not extend to the gas disk mass, for which no significant correlation with FUV flux ($F_{\mathrm{UV}}=1$–100\,G$_0$) is established. Even when dividing the sample into disks around low-mass stars with $M_{\star}< 0.4\,$M$_{\odot}$ and around high-mass stars with $M_{\star}> 0.4$\,M$_{\odot}$, no significant correlation is observed in both groups, as illustrated in Figure \ref{Fig:corr_mgas_fuv}. This is different from previous results for disks in the $\sigma$ Orionis star-forming region, where the strong FUV radiation field ($F_{\mathrm{UV}} = 10^2$–10$^5$\,G$_0$) leads to a pronounced decrease in dust disk mass with increasing FUV strength among high-mass stars, while disks around low-mass stars show little variation \citep{Mauco2023}. We emphasize that the results from \cite{Mauco2023} are for the dust disk mass and where disks are exposed to a stronger FUV radiation field that has likely activated only recently \citep{Coleman2025}. The current results from the AGE-PRO Large Program using the gas disk mass show no correlation between the gas disk mass and the FUV flux \citep{anania2025}, as predicted by our models.

The stellar mass and FUV flux remain essentially uncorrelated throughout the 10\,Myr evolution. Although the correlation coefficient r increases from $\sim 0.08$ at birth to $\sim 0.16$ at 10\,Myr (as shown in Table \ref{Tab:correlation}), the dependence is still very weak.

We compare each of our empirical correlations with AGE-PRO observations of 10 disks in Upper Sco, for which AGE-PRO provided gas disk masses and sizes \citep{Agurto-Gangas2025}. The left panel of Figure \ref{Fig:corr_mgas} suggests that four targets of AGE-PRO, Upper Sco 1, 3, 8 and 10 align well with the correlation between gas disk radius and mass. Upper Sco 9 has a larger gas disk mass, whereas the others have a lower mass than expected. Upper Sco 5 also exhibits a lower gas disk mass (right panel of Figure \ref{Fig:corr_mgas}) and a smaller gas disk radius (left panel of Figure \ref{Fig:corr_rco}) than expected for its stellar mass, possibly related to being a binary \citep{Trapman2025_5}. Upper Sco 6 also has a  smaller gas disk mass relative to the stellar mass. This may result from its cleared gas-cavity \citep{Vioque2025}.

The right panel of Figure \ref{Fig:corr_rco} suggests that Upper Sco 1, 6, 7, 8, 9 have larger gas disk radius compared to the experienced FUV flux. It is worth noting that all of these disks manifest inner dust cavities \citep{Vioque2025}. Although Upper Sco 10 also exhibits substructures, it does not exhibit strong variation from a smooth disk. This indicates that disk substructures, which are neglected in our simulations, may play important roles in gas disk evolution.

Furthermore, Upper Sco 1, 3 and 10 (marked in purple in Figure \ref{Fig:corr_mgas} and \ref{Fig:corr_rco}) exhibit hints of MHD winds from the presence of detected [O I] $\lambda$ 6300 line emission \citep{Fang2023}. These disks still largely follow the empirical correlations. Upper Sco 1 and 10 align well with the correlation between stellar mass, gas disk mass, and radius, whereas Upper Sco 3 exhibits systematically lower gas disk mass and radius for its stellar mass and the local FUV flux. Nevertheless, the gas disk mass of Upper Sco 3 still falls within the typical range for its stellar mass and radius at the given FUV flux. As Upper Sco 1 and 10 exhibit dust substructures, their relatively large gas disk radii at the corresponding FUV flux are likely driven by these substructures in the presence of MHD winds.

We note that the targets in AGE-PRO have stellar masses between 0.1 and 0.6 M$_{\odot}$, so we also restrict the results of the models to this range. The results for this restricted range of stellar mass are summarized in Figure \ref{Fig:corr_rco_1_6}. This gives a steeper slope for the correlations between gas disk radius and mass, FUV strength and gas disk radius. But still only four targets align well with our correlation between gas disk size and mass.

Moreover, we check the possible correlations between stellar mass and initial disk conditions. Gas disk evolution gives rise to a tentative positive correlation ($r \sim 0.5$) between the stellar mass and the viscosity parameter ($\alpha \propto M_{\star}^{0.65}$) among the surviving disks, and a weak positive correlation ($r \sim 0.2$) between the stellar mass and the characteristic radius ($R_c \propto M_{\star}^{0.17}$), with details shown in Table \ref{Tab:correlation}. For low-mass stars, disks with higher viscosity and larger characteristic radii are more efficiently dispersed by external photoevaporation, giving rise to the observed correlations between stellar mass, viscosity, and characteristic radius among the surviving disks.

Since the IMF yields only a small number of high-mass stars in a randomly drawn sample, we also verify our results by drawing stellar masses from a uniform distribution while keeping all other parameters the same. Based on this modified setup, we perform an additional population synthesis with 5000 samples. Similar results are obtained, as summarized in Table \ref{Tab:correlation_uniform}. However, adopting a uniform stellar mass distribution gives slightly different correlations. For example, the slope between the gas disk size and gas disk mass in log-log space at 5\,Myr is $\sim 22\%$ less than that in the IMF case. In addition, the correlation between stellar mass and FUV strength remains negligible at 10\,Myr, although its steady increase suggests that it may become significant at later evolutionary stages.

\section{Discussion}\label{Sec:discussion}
\subsection{A delayed effect of external photoevaporation}\label{subsec:delayed_effect}
The FUV flux that protoplanetary disks are exposed to evolves over time, mainly driven by the evolution in FUV luminosity of the massive stars and cluster dynamics. \citet{Kunitomo2021} emphasize the critical role of stellar evolution in shaping the temporal variation of photoevaporation rates. For stars with masses between 0.5 and 5 M$_{\odot}$, the FUV, EUV, and X-ray luminosities can vary by orders of magnitude within a few million years, leading to corresponding changes in the photoevaporation rates with time. Meanwhile, cluster dynamics change the relative distance between O,B stars and disks, thereby altering the FUV flux experienced by disks over time \citep{Concha-Ramirez2019,Concha-Ramirez2023}. Disks that are currently highly irradiated may have migrated from regions of lower irradiation, and conversely, some weakly irradiated disks may have originated in more strongly irradiated environments. The derived FUV flux from observations is the average of the FUV flux experienced by the disk, but it remains unknown what was the FUV flux experienced at earlier times. In addition, the disk may be shielded by the dust at early stages. N-body simulations indicate that dust extinction can efficiently shield disks in the first $<1$\,Myr \citep{Ali2019,Qiao2022,Wilhelm2023}.

\begin{figure}[!htbp]
  \centering
  \includegraphics[width=0.47\textwidth]{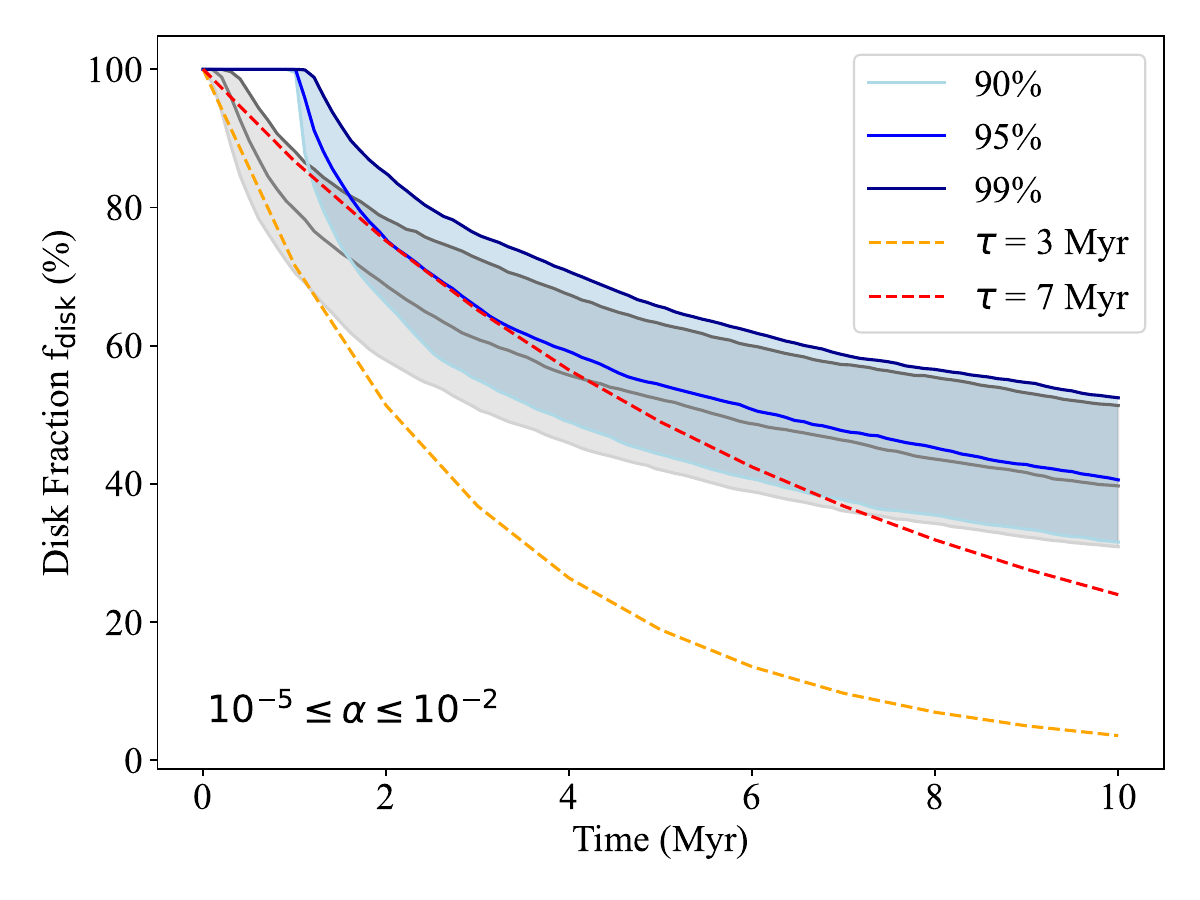} 
  \caption{The decline in disk fraction $f_{\mathrm{disk}}$ with the disk age. Same layout as Figure \ref{Fig:disk_fraction}. The gray and blue correspond to turning on external photoevaporation initially and at 1\,Myr. }
  \label{Fig:disk_fraction_1}
\end{figure}

To investigate the delayed effect of external photoevaporation, we perform a test where we turn on the external FUV flux at 1\,Myr in all our models.  No obvious differences in disk properties are observed shortly after the onset of external photoevaporation. The disks quickly evolve toward the same mass and radius as those predicted by models where photoevaporation is active from the beginning. The primary difference lies in the disk fraction, see Figure \ref{Fig:disk_fraction_1}. Prior to 1\,Myr, no disks are dispersed, as the disks in our models are not dispersed until $\sim 3$\,Myr under purely viscous evolution. After the activation of external photoevaporation, the evolution of the disk fraction follows the same trend as in models with continuous exposure to external FUV radiation from the outset. Nevertheless, at a given evolutionary stage, the disk fraction is systematically higher compared to models in which external photoevaporation is initiated at the onset. 

\subsection{Survivorship bias}
Figure \ref{Fig:gas_disk_mass} shows that the gas disk mass of surviving disks exhibits only limited evolution after 0.2\,Myr, suggesting survivorship bias. Such survivorship bias is also evident in the case of internal photoevaporation: inside-out dispersal significantly raises the median mass of the surviving population by rapidly removing low-mass disks, while outside-in dispersal of compact and light disks ($R_c \sim 10$\,AU) results in an almost constant median mass over time \citep{malanga2025}. In our scenario, external radiation keeps removing disk materials from the outer edge. Although low-mass disks disperse faster, potentially raising the median mass of survivors, this effect is counterbalanced by the overall mass loss from the population. As a result, the gas mass shows little evolution after 0.2\,Myr.

A similar survivorship bias is also observed in the gas disk size, which correlates with mass. Disks truncated to small radii by external photoevaporation are rapidly dispersed. As a result, disk dispersal offsets the shrinkage of most disks, keeping the median gas disk radius nearly constant after the first $\sim$ 0.2\,Myr. At later times $\sim$ 9 \,Myr, the median gas disk radius increases,  because the surviving disks are those able to sustain long-term viscous spreading despite ongoing external photoevaporation.

\subsection{Constraints on disk properties}\label{sub_diss:constrain_disk}
The comparison of our models with the observed disk fraction and gas disk size distribution predicts different ranges for the disk viscosity $\alpha$. Figure \ref{Fig:disk_fraction} favors a higher viscosity ($10^{-4} \leq \alpha \leq 10^{-2}$) as it leads to rapid disk dispersal. In contrast, Figure \ref{Fig:disk_Rco} indicates that a lower viscosity ($10^{-5} \leq \alpha \leq 10^{-2}$) is required to produce a large population of compact disks ($R_{^{12}\mathrm{CO},\,90\%} < 36$\,AU), more consistent with the observations. In this section, we discuss possible scenarios to reconcile these two ranges.

Building on the different preferences for viscosity, we further examine the possible dependence on the initial disk characteristic radius $R_{\mathrm{c}}$. For viscosity $10^{-5} \leq \alpha \leq 10^{-2}$, initially extended disks ($150 \leq R_{\mathrm{c}} \leq 200$\,AU) can reproduce the observed fractions but evolve into larger disks, whereas for $10^{-4} \leq \alpha \leq 10^{-2}$, initially compact disks ($10 \leq R_{\mathrm{c}} \leq 30$\,AU) can increase the fraction of compact disks with $R_{^{12}\mathrm{CO},\,90\%} < 36$\,AU, but disperse more slowly, as demonstrated in Figure \ref{Fig:disk_fraction_150_200} and \ref{Fig:rco_comparison_try}. The initial disk-to-star mass ratio $\epsilon$ also affects both the disk fraction and gas disk size distribution, but its influence is weaker than that of $R_{\mathrm{c}}$ and far less significant than that of viscosity. These results indicate that varying the constraints on viscosity parameter and characteristic radius alone cannot fully account for the observed trends. The dependence of disk fraction and gas disk size distribution on viscosity and characteristic radius also indicates a possible degeneracy between $R_{\mathrm{c}}$ and $\alpha$: initially compact disks with high viscosity and extended disks with low viscosity can produce similar outcomes in both disk fraction and gas disk size. \citet{Coleman2026} constrained the viscosity of a disk in Orion (d203-504) by combining the stellar accretion rate, external photoevaporative mass loss rate, stellar mass, and disk size and mass, finding $3\times10^{-4} \leq \alpha \leq 2\times10^{-3}$. This viscosity range fails to reproduce both the rapid dispersal of disks and the observed large population of compact disks. The initial conditions in Upper Sco may differ from those in Orion, making a direct comparison between the two regions non-trivial. Additional constraints on the wind mass loss rate and mass accretion rate may provide tighter constraints on disk properties \citep{Winter2020,Weder2026}.

\begin{figure}[!htbp]
  \centering
  \includegraphics[width=0.48\textwidth]{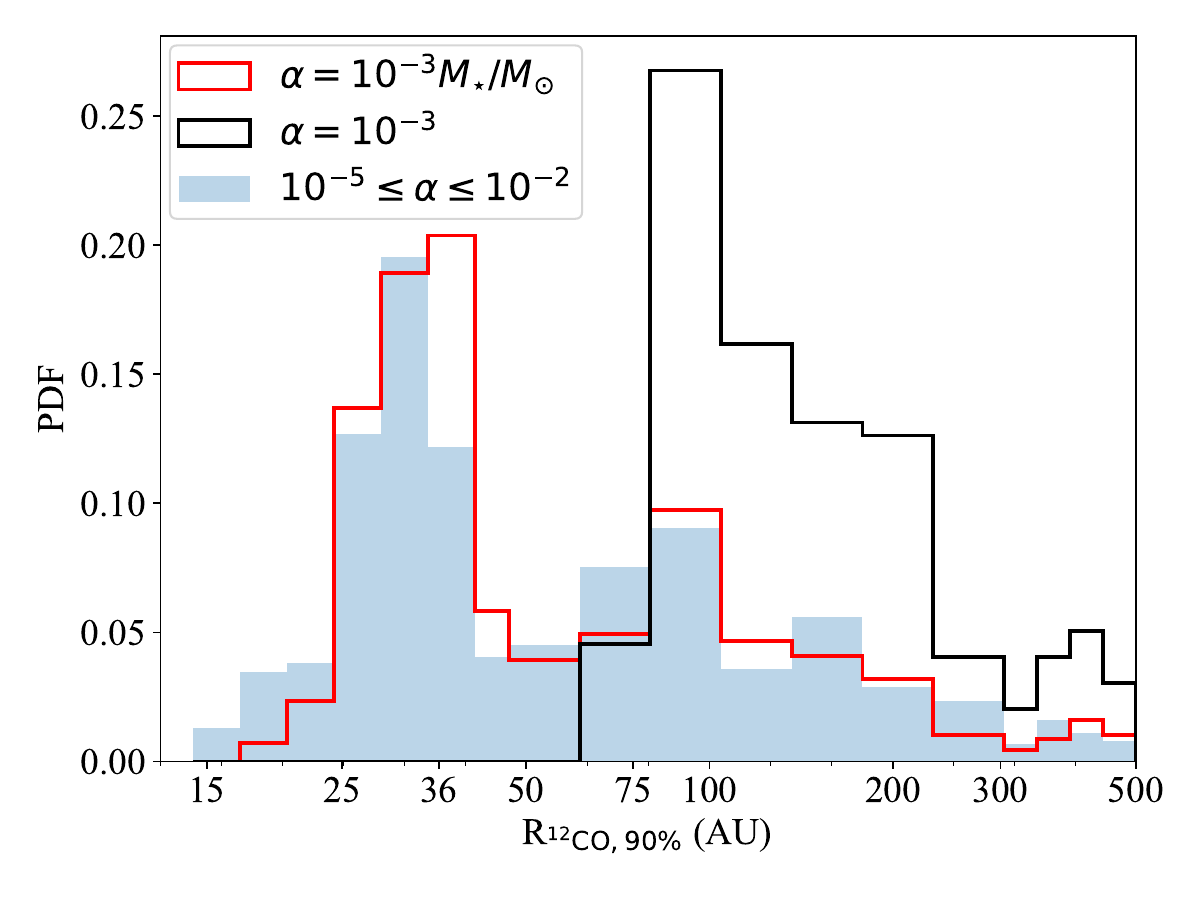} 
  \caption{The PDF of the gas disk radius $R_{^{12}\mathrm{CO},\,90\%}$ of surviving disks at 5\,Myr, assuming a constant viscosity $\alpha = 10^{-3}$ (black) and a mass-dependent prescription $\alpha = 10^{-3}M_{\star}/\mathrm{M_{\odot}}$ (red). The blue represents the whole population synthesis sample with viscosity $10^{-5} \leq \alpha \leq 10^{-2}$.}
  \label{Fig:rco_test_alpha}
\end{figure}

A promising way to reconcile the two viscosity ranges may lie in the potential correlation between the mass accretion rate and stellar mass. Observations reveal a strong correlation between stellar mass and accretion rate, scaling as $\dot{M}_{\mathrm{acc}} \propto M_{\star}^2$. This trend can be reproduced by disk evolution models that include a relation between viscosity parameter $\alpha$ and the stellar mass, such as $\alpha \propto M_{\star}$ \citep{Alexander2006}. Figure \ref{Fig:rco_test_alpha} compares the PDF of gas disk sizes for the constant viscosity $\alpha = 10^{-3}$ (black) and mass-dependent viscosity $\alpha = 10^{-3}M_{\star}/\mathrm{M_{\odot}}$ (red). The mass-dependent scenario effectively reduces the viscosity for disks around low-mass stars, producing more small disks. Conversely, with constant viscosity, the minimum gas disk size among our samples is $\sim 77$\,AU, corresponding to the disk around a $0.3$\,M$_{\odot}$ star. With a viscosity of $\alpha = 10^{-3}$, all disks around $0.1$\,M$_{\odot}$ stars are dispersed by 5\,Myr. These are also the ones that tend to have smaller radii, inferred from the strong correlation between stellar mass and gas disk radius in Sect. \ref{subsec:expected_corr}. This suggests that the commonly adopted assumption of $\alpha = 10^{-3}$ in viscous evolution models may not be suitable, or that additional mechanisms are required to ensure the survival of disks around low-mass stars under such viscosity. Those disks around low-mass stars, in particular, deserve special attention, as they are generally smaller and dominate the overall evolution of disk fractions given the large number of low-mass stars predicted by the IMF.

Although higher viscosity in disks around more massive stars can accelerate their dispersal, our test with $\alpha = 10^{-3} M_{\star}$/M$_{\odot}$ still leads to relatively slow evolution. A more detailed study of viscosity and its possible dependence on stellar mass is therefore needed to better explain both the observed disk fraction and distribution of gas disk sizes.

Furthermore, additional processes such as internal photoevaporation or MHD winds which are neglected in the current simulations, may accelerate disk dispersal, as discussed in Sect. \ref{sec_dis:add_mechanism}. 

\subsection{Disk substructures}
Our comparison of the empirical correlations from our population synthesis models with the AGE-PRO results in Upper Sco showed that disk substructures may play an important role in disk evolution. Disk substructures can be the result of dust trapping in pressure maxima, where grains reduce or stop their radial drift \citep{pinilla2012}. \cite{Garate2024} studied the survival and effect of dust traps on the evolution of disks that are exposed to external photoevaporation with $F_{\mathrm{UV}}$ between $10^{2}\text{-}10^{4}\,\mathrm{G_0}$. They found that if dust traps are located inside the photoevaporation truncation radius and  $F_{\mathrm{UV}}\leq10^{3}\,\mathrm{G_0}$, the presence of dust traps can extend the lifetime of the dust component. Moreover, the dust trap can alter the grain size distribution within the disk, in particular fragmentation of dust grains within pressure bumps can lead to a constant replenishment of small grains that can be easily entrained in the photoevaporative wind, possibly providing shielding against $F_{\mathrm{UV}}$ radiation since early times of evolution ($<0.1\,$Myr). This shielding may reduce the efficiency of external photoevaporation, allowing the gas disk to maintain a large size and evolve similarly to the purely viscous case. Consequently, our models that neglect dust substructures in the disk evolution may underestimate both the gas disk radius and its mass. The right panel of Figure \ref{Fig:corr_rco} suggests that this is the case, where Upper Sco 1, 6, 7, 8, 9, 10, which are disks with substructures \citep{Vioque2025} that exhibit larger gas disk radii compared to the correlation from our simulations. The inclusion of structures and dust evolution in population synthesis models that include external photoevaporation is important for understanding how substructures may change the disk lifetime and the properties of the surviving disks. 

\subsection{Additional processes affecting disk evolution}\label{sec_dis:add_mechanism}
Additional mechanisms neglected in our model, such as internal photoevaporation and MHD winds, can also affect disk evolution. Their potential impacts are briefly discussed in this section.

\paragraph{Internal Photoevaporation}
X-ray and FUV radiation from the central star can trigger a thermal wind and remove disk material. Generally, a gap will open when the photoevaporation rate and accretion rate become comparable, typically after a few Myr. Then the inner disk drains on the central star at its local viscous timescale, clearing the outer disk rapidly within $\sim 1$\,Myr \citep{Clarke2001,Picogna2019}. This process reduces the fraction of surviving disks. \citet{Coleman2022} demonstrated that when the external photoevaporative mass loss rate is below $\sim 10^{-8}$\,M$_\odot$\,yr$^{-1}$ ($F_{\mathrm{UV}} \sim$ 1000\,G$_0$), internal processes can significantly influence disk evolution. The extent of this impact depends on the radiation from the central star, as well as the disk viscosity and stellar mass, with internal photoevaporation becoming increasingly dominant for disks around higher-mass stars and at higher viscosities. As disks around high-mass stars also tend to have larger radii, internal photoevaporation may reduce the population of large gas disks with radius $R_{^{12}\mathrm{CO},\,90\%} > 150$\,AU. 

\citet{Tabone2025} modeled viscous disk evolution including internal photoevaporation but without external photoevaporation, and found that initially compact ($R_c \sim 5$–20\,AU) disks with low mass loss rates and relatively long viscous timescales ($\alpha \sim 2$–$4\times10^{-4}$) successfully reproduce the average gas disk sizes observed in Lupus, but not those in Upper Sco from AGE-PRO Large Program, and overestimate the apparent lifetime ($M/\dot{M}$). External photoevaporation can help to reduce this apparent lifetime  particularly in strong FUV environments. Therefore, models combining both internal and external photoevaporation may help in explaining the observed distributions.

\paragraph{MHD Winds}
Apart from internal photoevaporation, MHD winds launched from the disk surface along magnetic field lines can also influence the disk evolution. These winds are common and can contribute significantly to the disk mass loss. \citet{Fang2023} revealed the presence of MHD winds by the detection of emission of the [O I] $\lambda$ 6300 line in several Upper Sco disks. However, among the AGE-PRO targets with detected MHD winds (Upper Sco 1, 3, and 10), only Upper Sco 3 has lower gas disk mass and radius at the given stellar mass and the experienced FUV flux, illustrated in Figure \ref{Fig:corr_mgas} and \ref{Fig:corr_rco}. Notably, the other disks 1 and 10 exhibit substructures. This may suggest that, even in the presence of MHD winds and external photoevaporation, substructures can act to mitigate the loss of disk materials.

Our simulations indicate that the median gas disk mass of the surviving disks remains nearly constant after 0.2\,Myr, reflecting survivorship bias. MHD winds are predicted not to increase the disk radius as the disk evolves \citep{yang2021,Trapman2022}. Although viscously evolving disks tend to have slightly larger gas disk radii than MHD wind–driven disks at late evolutionary stages $\sim$ 10\,Myr, the difference is small \citep{Coleman2024,Weder2026,Pichierri2026}. Moreover, the gas disk radius exhibits large scatter due to the combination of different initial disk parameters (Figure \ref{Fig:corr_mgas} and \ref{Fig:corr_rco}). Therefore, under weak external photoevaporation levels of $\sim$ 10 \,G$_0$, the gas disk size alone cannot serve as a robust diagnostic to distinguish between viscous evolution and MHD winds \citep{Coleman2024,Weder2026}.

\citet{Tabone2022} demonstrated that in the MHD wind scenario, the median mass undergoes a steady decrease as the disk population evolves, resulting from ongoing accretion-driven depletion. Therefore, the time evolution of gas disk mass may offer a diagnostic to distinguish between different disk evolution scenarios: inside-out viscous evolution, which leads to an obvious increase in the median mass of the surviving disks; outside-in dispersal, where the median mass remains nearly constant with slight increase; and MHD winds, which instead drive a decrease in median gas disk mass. However, since \citet{Tabone2022} does not account for external photoevaporation, the decreasing trend of the median gas disk mass predicted by MHD wind-driven models may be altered, and needs to be investigated in the future.

More complex models with disk substructures, internal photoevaporation and MHD winds are necessary to better understand disk evolution.

\section{Conclusions} \label{Sec:conclusions}
In this work, we perform population synthesis models of the gas evolution of disks exposed to mild values of external FUV flux ($F_{\mathrm{UV}} =1$–100\,G$_0$), and explore the following initial conditions: stellar mass distributed as the IMF, disk viscosity ($\alpha$), disk mass, and characteristic disk radius ($R_\mathrm{c}$). We aim to constrain the properties of the disks that survive up to 5–10\,Myr of evolution, and determine how the disk fraction changes over time. We compare our stellar mass distribution and gas disk size distribution with the sample of \citet{carpenter2025}, as well as with the gas disk masses and sizes of the 10 targets in Upper Sco from the AGE-PRO Large Program. Our conclusions are as follows.
\begin{itemize}
    \item Observations referring to the disk fraction in star-forming regions of different age yield disk lifetimes of $\sim$ 3–7\,Myr. Our population synthesis can explain the observed exponential trend, particularly for viscosities in the range $10^{-4}\leq\alpha\leq10^{-2}$ (Fig.~\ref{Fig:disk_fraction}).

    \item The percentage of surviving disks around stars of mass 0.1\,M$_\odot$ decreases over time, as the gas in disks around less massive stars is less gravitationally bound, hence easier to be entrained in photoevaporative winds. Initially, the percentage of stars in the sample around 0.1\, M$_\odot$ is 76\%, and it is only around 51\% after 10\,Myr of evolution, which is about 21\% lower than what is observed in the Upper Sco region. This significant decrease in the number of surviving disks is not found for more massive stars (Fig.~\ref{Fig:survive_mstar}).

    \item The majority of disks that survived 5–10\,Myr require low viscosity ($\alpha<10^{-3.5}$) and low characteristic radius ($R_\mathrm{c}<125$\,AU), as a low viscosity and a small $R_\mathrm{c}$ result in more compact disks that are more gravitationally bound and less vulnerable to mass loss by external photoevaporation (Fig.~\ref{Fig:survive_alpha_rc}). In addition, surviving disks do not show a significant preference for higher initial disk-to-star mass ratios.
    
    \item The median gas disk mass of the surviving disks decreases sharply within the first 0.2\,Myr of evolution, and increases slightly thereafter, reflecting survivorship bias. 
    
    \item The gas disk size distribution of the disk population also exhibits survivorship bias, with its median value dropping sharply after 0.2\,Myr and remaining nearly constant throughout the subsequent evolution. Higher viscosity $10^{-4}\leq\alpha\leq 10^{-2}$ fails to produce the large population of compact disks found in the sample of \citet{carpenter2025} in Upper Sco (Fig.~\ref{Fig:disk_Rco}).
    
    \item We investigated potential correlations between observable quantities. We find a moderate correlation ($r \sim 0.6$) between $M_{\mathrm{gas}}$ and $R_{^{12}\mathrm{CO},\,90\%}$ ($M_{\mathrm{gas}} \propto R_{^{12}\mathrm{CO},\,90\%}^{1.58}$), and a tentative correlation ($r \sim 0.5$) between $M_{\mathrm{gas}}$ and stellar mass $M_\star$ ($M_{\mathrm{gas}} \propto M_{\star}^{1.29}$). The correlation between $M_{\mathrm {gas}}$ and $F_{\mathrm UV}$ is negligible ($r \sim 0.04$), consistent with the findings of the AGE-PRO collaboration \citep{anania2025}. Moreover, there is a strong correlation ($r \sim 0.9$) between $R_{^{12}\mathrm{CO},\,90\%}$ and $M_{\star}$ ($R_{^{12}\mathrm{CO},\,90\%} \propto M_{\star}^{0.87}$), and a very weak negative correlation ($r \sim -0.2$) between $R_{^{12}\mathrm{CO},\,90\%}$ and FUV strength ($R_{^{12}\mathrm{CO},\,90\%} \propto F_{\mathrm{UV}}^{-0.14}$) (Figs.~\ref{Fig:corr_mgas} and~\ref{Fig:corr_rco}).
\end{itemize}

This work highlights the importance of the process of external photoevaporation at a disk population level even in the presence of mild levels of FUV radiation field ($F_{\mathrm{UV}}=1$–100\,G$_0$), as in the case of Upper Sco. To constrain the models and the disk parameters, further observations of gas disk properties, especially mass and size are needed. Moreover, as disks with substructures exhibit larger gas disk radii than expected for a given FUV flux (Fig. \ref{Fig:corr_rco}), our results emphasize that additional mechanisms, in particular disk substructures, should be incorporated into population synthesis models.

\begin{acknowledgements}
We thank the referee for useful comments and suggestions that helped improve the manuscript's quality. RA acknowledges funding from the Fondazione Cariplo, grant no. 2022-1217, and the European Research Council (ERC) under the European Union’s Horizon Europe Research \& Innovation Programme under grant agreement no. 101039651 (DiscEvol). Views and opinions expressed are however those of the author(s) only, and do not necessarily reflect those of the European Union or the European Research Council Executive Agency. Neither the European Union nor the granting authority can be held responsible for them.
PP acknowledges funding from the UK Research and Innovation (UKRI) under the UK government’s Horizon Europe funding guarantee from ERC (under grant agreement No 101076489).
\end{acknowledgements}

\bibliographystyle{aa} 
\bibliography{ref}

\begin{appendix}

\section{Variable luminosity} \label{app:variable_l}
In this appendix, we illustrate the time evolution of disk fraction when the luminosity of the central star evolves over time.

\begin{figure}[!htbp]
  \centering
  \includegraphics[width=0.47\textwidth]{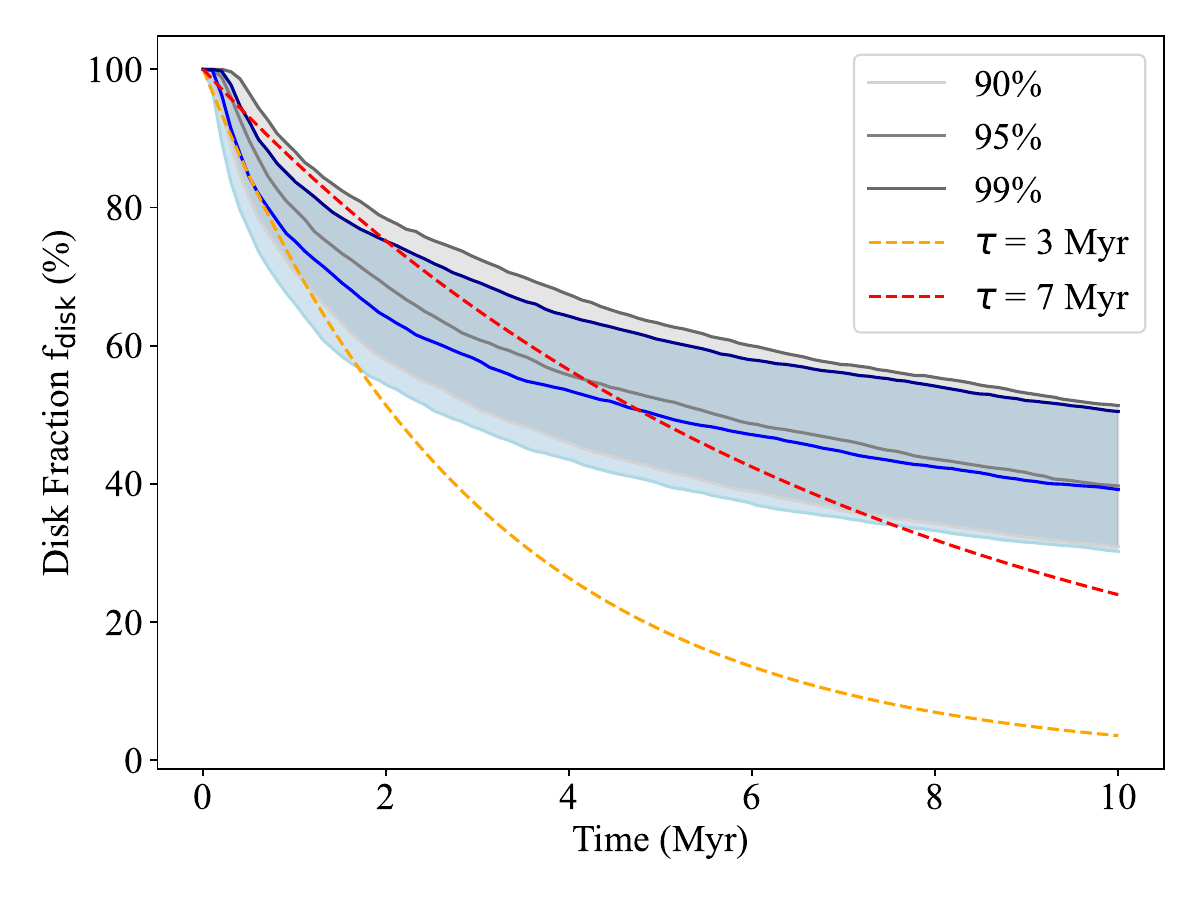} 
  \caption{The decline in disk fraction $f_{\mathrm{disk}}$ with the disk age. Similar layout to Figure \ref{Fig:disk_fraction}. The gray and blue represent constant luminosity for stars at the age of 5\,Myr and luminosity that evolves following the evolutionary track by \citet{Dotter2008}.  }
  \label{Fig:disk_frac_vari_L}
\end{figure}

\section{Disk fraction} \label{app:disk_frac}
In this appendix, we demonstrate the disk fraction considering different ranges of viscosity $\alpha$. 

\begin{figure}[t!]
  \centering
    \includegraphics[width=0.88\linewidth]{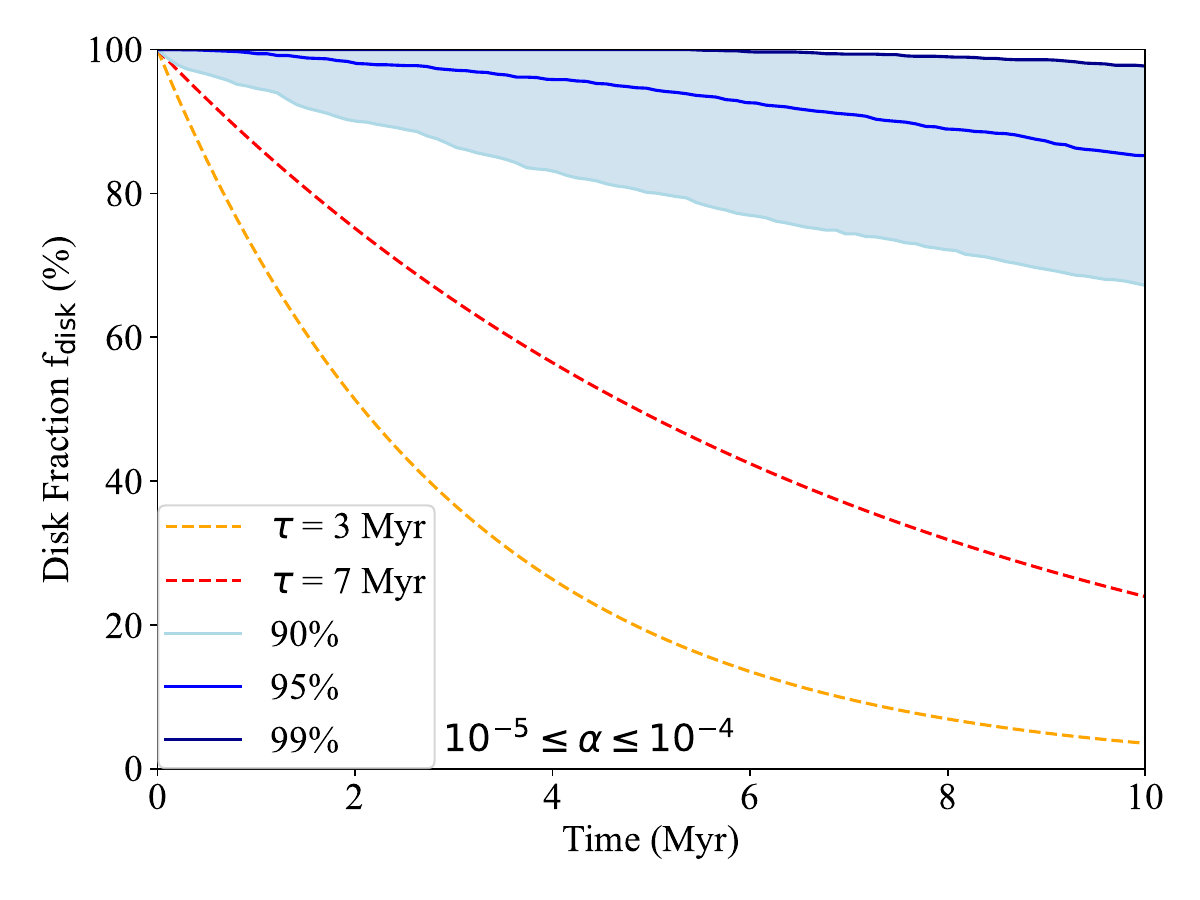}
    \includegraphics[width=0.88\linewidth]{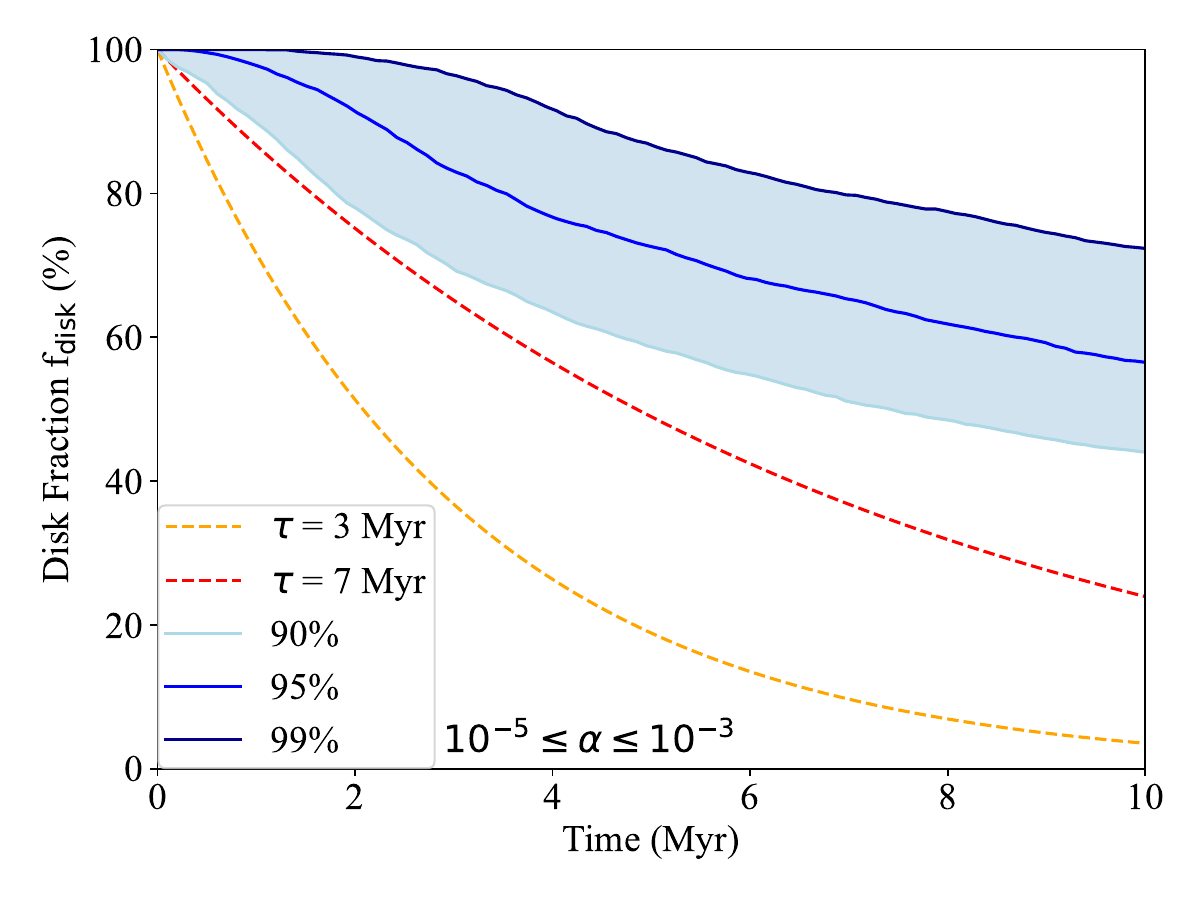}
   
    \includegraphics[width=0.88\linewidth]{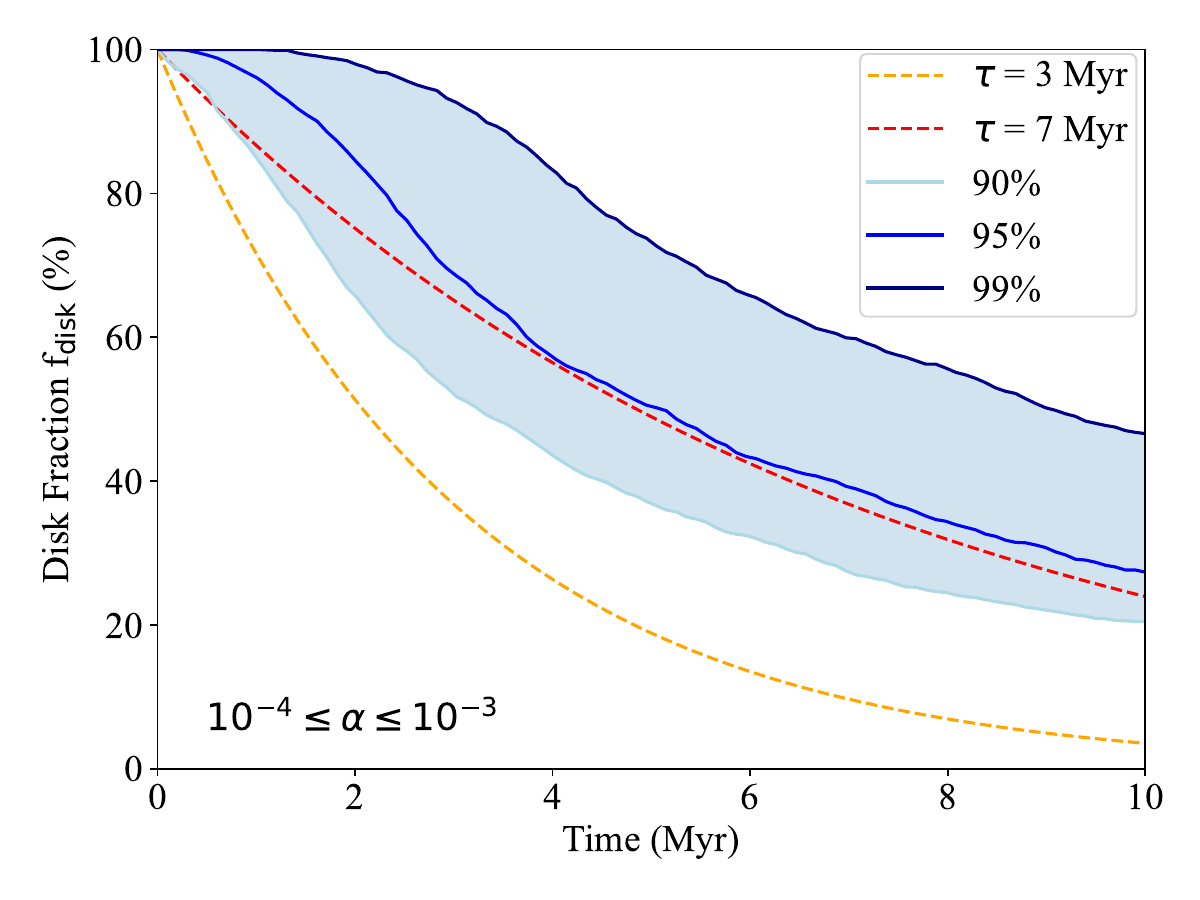}
    \includegraphics[width=0.88\linewidth]{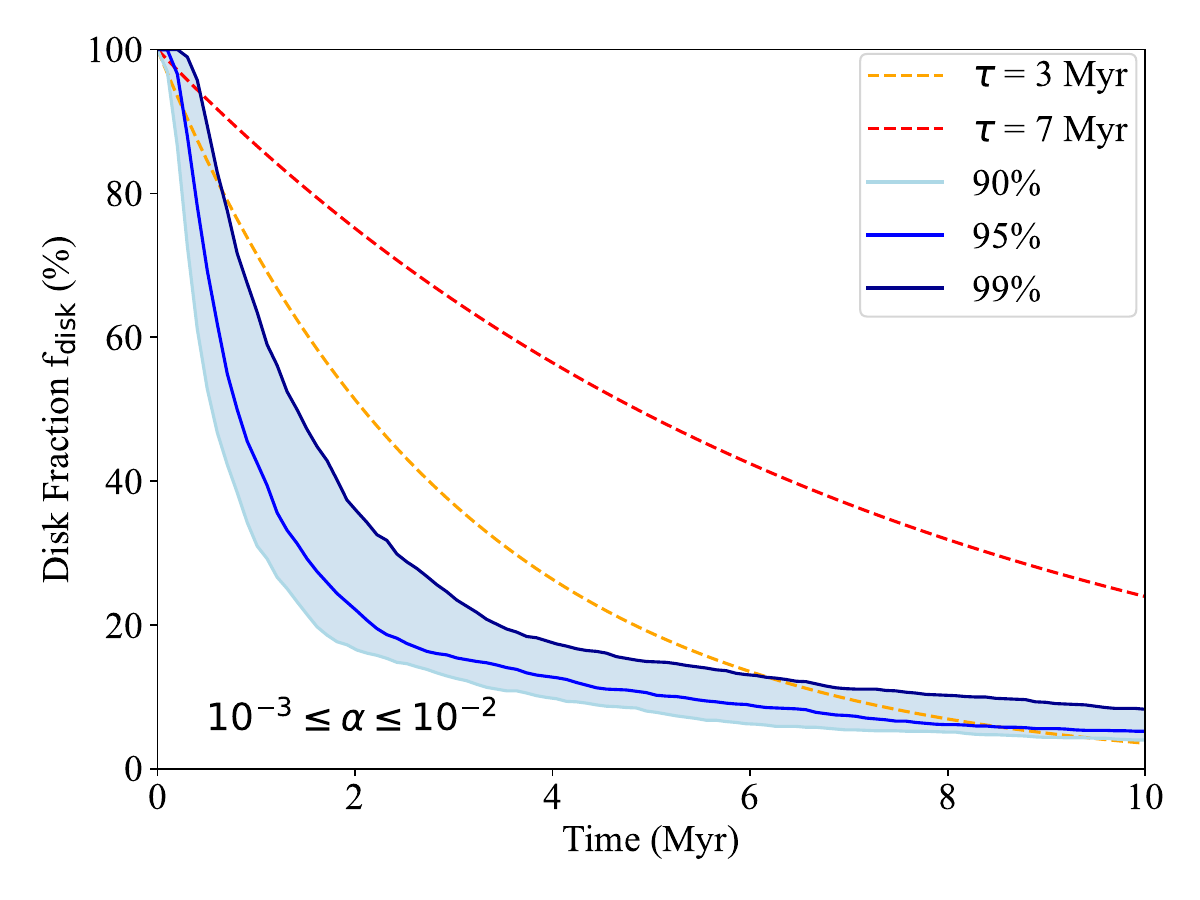}

    \captionof{figure}{Disk fraction with the disk age limited to different ranges of viscosity $\alpha$. Same layout as Figure \ref{Fig:disk_fraction}.}
    \label{Fig:app_diskfrac}

\end{figure}

\clearpage
\section{Empirical correlations}\label{app:tab_imf}
\begin{figure}[htbp]
  \centering
  \includegraphics[width=0.97\linewidth]{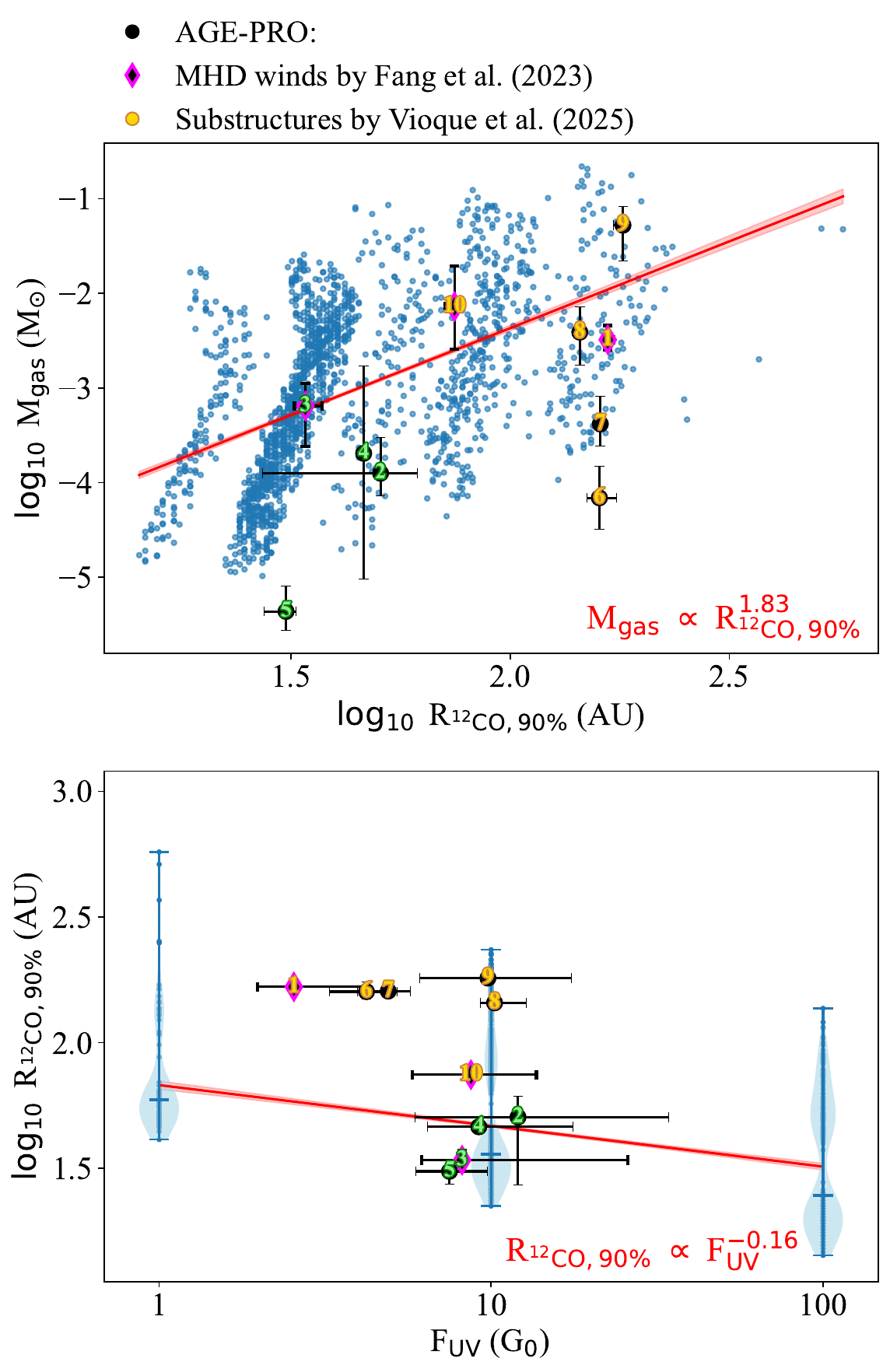}
  \caption{Empirical correlations between the gas disk mass and gas disk radius (top), the FUV flux and gas disk radius (bottom) at 5\,Myr, restricted to disks around 0.1–0.6\,M$_{\odot}$ stars. Same layout as Figure \ref{Fig:corr_mgas} and \ref{Fig:corr_rco}.}
  \label{Fig:corr_rco_1_6}
\end{figure}

In this appendix, we restrict our samples to the disks around 0.1–0.6\, M$_{\odot}$ stars to compare with the ten targets studied by the AGE-PRO ALMA Large Program. The resulting correlations between the gas disk radius and mass, the FUV flux and gas disk radius are demonstrated in Figure \ref{Fig:corr_rco_1_6}.

We also show the correlation between surviving disks and stellar properties at different times of disk evolution. Table \ref{Tab:correlation} demonstrates the empirical correlations with stellar mass drawn from the IMF, which is discussed in Sect. \ref{subsec:expected_corr}. And Table \ref{Tab:correlation_uniform} shows the empirical correlations similar to Table \ref{Tab:correlation}
but with stellar mass drawn from a uniform distribution to avoid the bias from small number of high-mass stars given by the IMF. Table \ref{Tab:correlation_uniform} gives similar results as Table \ref{Tab:correlation}, with differences in the slope and correlation coefficient. For example, uniform stellar mass distribution gives steeper correlation between gas disk radius and FUV strength, $\sim 42\%$ steeper at 5\,Myr. And for uniform stellar mass distribution, the slope remains nearly constant between 5 and 10\,Myr, while for the IMF, the slope decreases by $\sim 36\%$. 

\begin{table*}[!htbp]
  \centering
  \caption{Correlations between surviving disk and stellar properties at different ages of disk evolution considering the IMF.}
  \label{Tab:correlation}
  \renewcommand{\arraystretch}{1.6}
  \footnotesize
  \begin{tabular}{cccccc}
    \hline
    \hline
    x  & y  & Time (Myr) & r  & $\alpha$ & $\beta$   \\
    \hline   
    \multirow{4}{*}{$\log_{10} R_{^{12}\mathrm{CO}, \, 90\%}$\,(AU)}&\multirow{4}{*}{$\log_{10} M_{\mathrm{gas}}$\,(M$_{\odot}$)} & 1 & $0.601^{+0.012}_{-0.011}$ & $1.737^{+0.039}_{-0.038}$ & $-5.879^{+0.068}_{-0.068}$  \\
    & & 3&  $0.613^{+0.013}_{-0.013}$ & $1.596^{+0.042}_{-0.041}$ & $-5.644^{+0.072}_{-0.075}$\\
    & & 5 & $0.624^{+0.014}_{-0.013}$ & $1.576^{+0.042}_{-0.044}$ & $-5.622^{+0.078}_{-0.077}$\\
    & &  10&$0.624^{+0.016}_{-0.016}$ & $1.500^{+0.048}_{-0.047}$ & $-5.494^{+0.086}_{-0.088}$ \\
    \hline
    \multirow{4}{*}{$\log_{10} M_{\star}$ (M$_{\odot}$)} &\multirow{4}{*}{ $\log_{10} M_{\mathrm{gas}}$\,(M$_{\odot}$)}&1 & $0.524^{+0.012}_{-0.012}$ & $1.332^{+0.037}_{-0.035}$ & $-1.851^{+0.031}_{-0.031}$ \\
    & & 3 & $0.540^{+0.014}_{-0.014}$ & $1.285^{+0.039}_{-0.040}$ & $-1.941^{+0.032}_{-0.033}$ \\
    & & 5 & $0.553^{+0.015}_{-0.015}$ & $1.288^{+0.042}_{-0.042}$ & $-1.970^{+0.033}_{-0.034}$ \\
    & & 10 & $0.560^{+0.017}_{-0.017}$ & $1.239^{+0.047}_{-0.045}$ & $-2.022^{+0.035}_{-0.036}$ \\
    \hline
    \multirow{4}{*}{$\log_{10} M_{\star}$ (M$_{\odot}$)} &\multirow{4}{*}{ $\log_{10} \epsilon\,(=M_{\mathrm{gas}}/M_{\star})$}&1 & $0.152^{+0.016}_{-0.017}$ & $0.332^{+0.035}_{-0.036}$ & $-1.851^{+0.030}_{-0.032}$ \\
    & & 3 & $0.141^{+0.018}_{-0.019}$ & $0.285^{+0.038}_{-0.039}$ & $-1.941^{+0.031}_{-0.033}$ \\
    & & 5 &$0.147^{+0.022}_{-0.021}$ & $0.289^{+0.043}_{-0.043}$ & $-1.970^{+0.035}_{-0.033}$ \\
    & & 10 & $0.130^{+0.025}_{-0.025}$ & $0.240^{+0.047}_{-0.046}$ & $-2.022^{+0.036}_{-0.036}$ \\
    \hline
    \multirow{4}{*}{$\log_{10} F_{\mathrm{UV}}$ ($\mathrm{G_0}$)}&\multirow{4}{*}{ $\log_{10} M_{\mathrm{gas}}$\,(M$_{\odot}$)}&1 & $0.026^{+0.017}_{-0.017}$ & $0.055^{+0.035}_{-0.036}$ & $-2.940^{+0.045}_{-0.044}$ \\
    & &3 &$0.040^{+0.019}_{-0.019}$ & $0.087^{+0.043}_{-0.042}$ & $-2.955^{+0.053}_{-0.053}$ \\
    & & 5& $0.045^{+0.023}_{-0.022}$ & $0.102^{+0.051}_{-0.050}$ & $-2.962^{+0.061}_{-0.061}$ \\
    & & 10 &$0.100^{+0.025}_{-0.026}$ & $0.223^{+0.057}_{-0.058}$ & $-3.043^{+0.070}_{-0.067}$ \\
    \hline
    \multirow{4}{*}{$\log_{10} M_{\star}$ (M$_{\odot}$)}&\multirow{4}{*}{$\log_{10} R_{^{12}\mathrm{CO}, \, 90\%}$\,(AU)}&1 & $0.920^{+0.003}_{-0.003}$ & $0.810^{+0.005}_{-0.006}$ & $2.351^{+0.005}_{-0.005}$ \\
    & &3 &$0.934^{+0.002}_{-0.003}$ & $0.853^{+0.007}_{-0.006}$ & $2.355^{+0.005}_{-0.005}$ \\
    & & 5& $0.940^{+0.002}_{-0.003}$ & $0.867^{+0.007}_{-0.006}$ & $2.353^{+0.005}_{-0.006}$ \\
    & & 10 &$0.945^{+0.002}_{-0.003}$ & $0.871^{+0.008}_{-0.007}$ & $2.343^{+0.006}_{-0.006}$ \\
    \hline
    \multirow{4}{*}{$\log_{10} F_{\mathrm{UV}}$ ($\mathrm{G_0}$)}&\multirow{4}{*}{$\log_{10} R_{^{12}\mathrm{CO}, \, 90\%}$\,(AU)}&1 & $-0.208^{+0.016}_{-0.016}$ & $-0.153^{+0.012}_{-0.012}$ & $1.906^{+0.015}_{-0.015}$ \\
    & &3 &$-0.173^{+0.020}_{-0.019}$ & $-0.145^{+0.016}_{-0.016}$ & $1.914^{+0.020}_{-0.020}$ \\
    & & 5& $-0.151^{+0.022}_{-0.021}$ & $-0.135^{+0.019}_{-0.019}$ & $1.915^{+0.024}_{-0.023}$ \\
    & & 10 &$-0.093^{+0.025}_{-0.025}$ & $-0.087^{+0.024}_{-0.023}$ & $1.897^{+0.029}_{-0.028}$ \\
    \hline
    \multirow{4}{*}{$\log_{10} M_{\star}$ (M$_{\odot}$)} &\multirow{4}{*}{ $\log_{10} F_{\mathrm{UV}}$\,($\mathrm{G_0}$)}&1 & $0.079^{+0.016}_{-0.017}$ & $0.095^{+0.019}_{-0.020}$ & $1.236^{+0.017}_{-0.017}$ \\
    & & 3 & $0.096^{+0.019}_{-0.019}$ & $0.105^{+0.021}_{-0.021}$ & $1.218^{+0.017}_{-0.018}$ \\
    & & 5 &$0.108^{+0.022}_{-0.021}$ & $0.112^{+0.022}_{-0.022}$ & $1.211^{+0.018}_{-0.018}$ \\
    & & 10 & $0.157^{+0.025}_{-0.024}$ & $0.156^{+0.025}_{-0.025}$ & $1.213^{+0.019}_{-0.019}$ \\
    \hline
    \multirow{4}{*}{$\log_{10} M_{\star}$ (M$_{\odot}$)} &\multirow{4}{*}{ $\log_{10} R_{\mathrm{c}}$\,(AU)}
    &1 & $0.094^{+0.017}_{-0.016}$ & $0.089^{+0.015}_{-0.016}$ & $1.652^{+0.014}_{-0.013}$ \\
    & & 3 & $0.156^{+0.019}_{-0.020}$ & $0.130^{+0.017}_{-0.016}$ & $1.652^{+0.013}_{-0.014}$ \\
    & & 5 &$0.206^{+0.021}_{-0.021}$ & $0.165^{+0.017}_{-0.017}$ & $1.654^{+0.014}_{-0.013}$ \\
    & & 10 & $0.255^{+0.024}_{-0.024}$ & $0.195^{+0.019}_{-0.019}$ & $1.651^{+0.015}_{-0.014}$ \\
    \hline
    \multirow{4}{*}{$\log_{10} M_{\star}$ (M$_{\odot}$)} &\multirow{4}{*}{ $\log_{10} \alpha$}
    &1 & $0.246^{+0.015}_{-0.016}$ & $0.465^{+0.030}_{-0.031}$ & $-3.474^{+0.026}_{-0.026}$ \\
    & & 3 & $0.408^{+0.017}_{-0.017}$ & $0.638^{+0.028}_{-0.028}$ & $-3.587^{+0.023}_{-0.024}$ \\
    & & 5 &$0.461^{+0.016}_{-0.018}$ & $0.655^{+0.027}_{-0.027}$ & $-3.691^{+0.023}_{-0.021}$ \\
    & & 10 & $0.522^{+0.018}_{-0.018}$ & $0.653^{+0.027}_{-0.027}$ & $-3.848^{+0.020}_{-0.021}$ \\
    \hline
  \end{tabular}
  \tablefoot{Summary of the correlations and fits among different disk and stellar properties at 1, 3, 5 and 10\,Myr using \texttt{linmix} \citep{Kelly2007}. The fitting function is $y=\alpha x+ \beta$. And r represents the Pearson correlation coefficient. The median and 1$\sigma$ interval are listed. As we do not apply any uncertainty to our simulations, the resulting 1$\sigma$ credible intervals are correspondingly narrow.}
\end{table*}

\begin{table*}[!htbp]
  \centering
  \caption{Correlations between surviving disk and stellar properties at different ages of disk evolution considering the uniform stellar mass distribution, constructed in the same way as Table  \ref{Fig:corr_mgas}.}
  \label{Tab:correlation_uniform}
  \renewcommand{\arraystretch}{1.6}
  \footnotesize
  \begin{tabular}{cccccc}
    \hline
    \hline
    x  & y  & Time (Myr) & r  & $\alpha$ & $\beta$   \\
    \hline
    \multirow{4}{*}{$\log_{10} R_{^{12}\mathrm{CO}, \, 90\%}$\,(AU)}&\multirow{4}{*}{$\log_{10} M_{\mathrm{gas}}$\,(M$_{\odot}$)} 
    & 1 & $0.591^{+0.009}_{-0.009}$ & $1.469^{+0.029}_{-0.029}$ & $-5.306^{+0.067}_{-0.067}$  \\
    & & 3&  $0.542^{+0.010}_{-0.011}$ & $1.287^{+0.031}_{-0.031}$ & $-4.961^{+0.073}_{-0.071}$\\
    & & 5 & $0.512^{+0.011}_{-0.012}$ & $1.211^{+0.033}_{-0.034}$ & $-4.805^{+0.079}_{-0.078}$\\
    & &  10&$0.506^{+0.013}_{-0.012}$ & $1.202^{+0.036}_{-0.036}$ & $-4.794^{+0.086}_{-0.084}$ \\
    \hline
    \multirow{4}{*}{$\log_{10} M_{\star}$ (M$_{\odot}$)} &\multirow{4}{*}{ $\log_{10} M_{\mathrm{gas}}$\,(M$_{\odot}$)}
    &1 & $0.574^{+0.009}_{-0.010}$ & $1.279^{+0.026}_{-0.027}$ & $-1.830^{+0.012}_{-0.012}$ \\
    & & 3 & $0.532^{+0.011}_{-0.011}$ & $1.225^{+0.030}_{-0.030}$ & $-1.902^{+0.012}_{-0.013}$ \\
    & & 5 & $0.505^{+0.012}_{-0.012}$ & $1.186^{+0.034}_{-0.032}$ & $-1.929^{+0.013}_{-0.014}$ \\
    & & 10 & $0.493^{+0.013}_{-0.013}$ & $1.185^{+0.038}_{-0.037}$ & $-1.949^{+0.014}_{-0.014}$ \\
    \hline
    \multirow{4}{*}{$\log_{10} M_{\star}$\,(M$_{\odot}$)} &\multirow{4}{*}{ $\log_{10} \epsilon=M_{\mathrm{gas}}/M_{\star}$}
    &1 & $0.151^{+0.014}_{-0.014}$ & $0.278^{+0.027}_{-0.026}$ & $-1.830^{+0.013}_{-0.012}$ \\
    & & 3 & $0.115^{+0.015}_{-0.016}$ & $0.225^{+0.030}_{-0.030}$ & $-1.902^{+0.013}_{-0.013}$ \\
    & & 5 &$0.092^{+0.016}_{-0.016}$ & $0.187^{+0.033}_{-0.033}$ & $-1.929^{+0.013}_{-0.013}$ \\
    & & 10 & $0.088^{+0.018}_{-0.017}$ & $0.186^{+0.036}_{-0.037}$ & $-1.949^{+0.015}_{-0.014}$ \\
    \hline
    \multirow{4}{*}{$\log_{10} F_{\mathrm{UV}}$ ($\mathrm{G_0}$)}&\multirow{4}{*}{ $\log_{10} M_{\mathrm{gas}}$\,(M$_{\odot}$)}
    &1 & $0.038^{+0.015}_{-0.015}$ & $0.081^{+0.032}_{-0.032}$ & $-2.078^{+0.040}_{-0.040}$ \\
    & &3 &$0.029^{+0.015}_{-0.016}$ & $0.061^{+0.033}_{-0.033}$ & $-2.063^{+0.041}_{-0.040}$ \\
    & & 5& $0.043^{+0.016}_{-0.017}$ & $0.090^{+0.035}_{-0.034}$ & $-2.094^{+0.043}_{-0.042}$ \\
    & & 10 &$0.057^{+0.017}_{-0.018}$ & $0.119^{+0.037}_{-0.036}$ & $-2.116^{+0.045}_{-0.045}$ \\
    \hline
    \multirow{4}{*}{$\log_{10} M_{\star}$ (M$_{\odot}$)}&\multirow{4}{*}{$\log_{10} R_{^{12}\mathrm{CO}, \, 90\%}$\,(AU) }
    &1 & $0.875^{+0.003}_{-0.004}$ & $0.784^{+0.007}_{-0.006}$ & $2.357^{+0.003}_{-0.003}$ \\
    & &3 &$0.881^{+0.003}_{-0.004}$ & $0.853^{+0.007}_{-0.007}$ & $2.369^{+0.003}_{-0.004}$ \\
    & & 5& $0.881^{+0.003}_{-0.004}$ & $0.874^{+0.008}_{-0.008}$ & $2.370^{+0.003}_{-0.003}$ \\
    & & 10 &$0.887^{+0.003}_{-0.004}$ & $0.898^{+0.008}_{-0.008}$ & $2.364^{+0.004}_{-0.003}$ \\
    \hline
    \multirow{4}{*}{$\log_{10} F_{\mathrm{UV}}$ ($\mathrm{G_0}$)}&\multirow{4}{*}{$\log_{10} R_{^{12}\mathrm{CO}, \, 90\%}$\,(AU) }
    &1 & $-0.181^{+0.014}_{-0.014}$ & $-0.155^{+0.012}_{-0.013}$ & $2.445^{+0.015}_{-0.016}$ \\
    & &3 &$-0.215^{+0.015}_{-0.015}$ & $-0.192^{+0.014}_{-0.013}$ & $2.529^{+0.016}_{-0.017}$ \\
    & & 5& $-0.219^{+0.015}_{-0.016}$ & $-0.196^{+0.014}_{-0.014}$ & $2.549^{+0.018}_{-0.017}$ \\
    & & 10 &$-0.218^{+0.018}_{-0.017}$ & $-0.192^{+0.015}_{-0.015}$ & $2.560^{+0.018}_{-0.019}$ \\
    \hline
    \multirow{4}{*}{$\log_{10} M_{\star}$ (M$_{\odot}$)} &\multirow{4}{*}{ $\log_{10} F_{\mathrm{UV}}$\,($\mathrm{G_0}$)}
    &1 & $0.046^{+0.015}_{-0.014}$ & $0.049^{+0.015}_{-0.015}$ & $1.176^{+0.007}_{-0.007}$ \\
    & & 3 & $0.034^{+0.016}_{-0.015}$ & $0.037^{+0.017}_{-0.017}$ & $1.163^{+0.007}_{-0.007}$ \\
    & & 5 &$0.046^{+0.015}_{-0.017}$ & $0.051^{+0.017}_{-0.018}$ & $1.151^{+0.008}_{-0.007}$ \\
    & & 10 & $0.060^{+0.018}_{-0.017}$ & $0.069^{+0.020}_{-0.020}$ & $1.143^{+0.008}_{-0.008}$ \\
    \hline
    \multirow{4}{*}{$\log_{10} M_{\star}$ (M$_{\odot}$)} &\multirow{4}{*}{ $\log_{10} R_{\mathrm{c}}$\,(AU)}
    &1 & $0.054^{+0.015}_{-0.015}$ & $0.046^{+0.012}_{-0.013}$ & $1.638^{+0.006}_{-0.005}$ \\
    & & 3 & $0.070^{+0.015}_{-0.015}$ & $0.063^{+0.013}_{-0.014}$ & $1.636^{+0.006}_{-0.005}$ \\
    & & 5 &$0.083^{+0.016}_{-0.016}$ & $0.076^{+0.015}_{-0.015}$ & $1.635^{+0.007}_{-0.006}$ \\
    & & 10 & $0.102^{+0.017}_{-0.017}$ & $0.098^{+0.016}_{-0.017}$ & $1.634^{+0.006}_{-0.007}$ \\
    \hline
     \multirow{4}{*}{$\log_{10} M_{\star}$ (M$_{\odot}$)} &\multirow{4}{*}{ $\log_{10} \alpha$}
    &1 & $0.136^{+0.014}_{-0.015}$ & $0.261^{+0.028}_{-0.028}$ & $-3.545^{+0.012}_{-0.013}$ \\
    & & 3 & $0.240^{+0.014}_{-0.015}$ & $0.467^{+0.029}_{-0.030}$ & $-3.641^{+0.012}_{-0.013}$ \\
    & & 5 &$0.268^{+0.015}_{-0.015}$ & $0.514^{+0.030}_{-0.030}$ & $-3.731^{+0.012}_{-0.013}$ \\
    & & 10 & $0.288^{+0.016}_{-0.016}$ & $0.516^{+0.031}_{-0.029}$ & $-3.883^{+0.011}_{-0.012}$ \\
    \hline
  \end{tabular}
\end{table*}
\clearpage
\section{Dependence on characteristic radius}
In this appendix, we check how the disk fraction and gas disk size distribution depend on the characteristic radius, considering the two viscosity ranges inferred from the disk fraction ($10^{-4} \leq \alpha \leq 10^{-2}$) and gas disk size distribution ($10^{-5} \leq \alpha \leq 10^{-2}$).

\begin{figure}[htbp]
  \centering
  \includegraphics[width=0.47\textwidth]{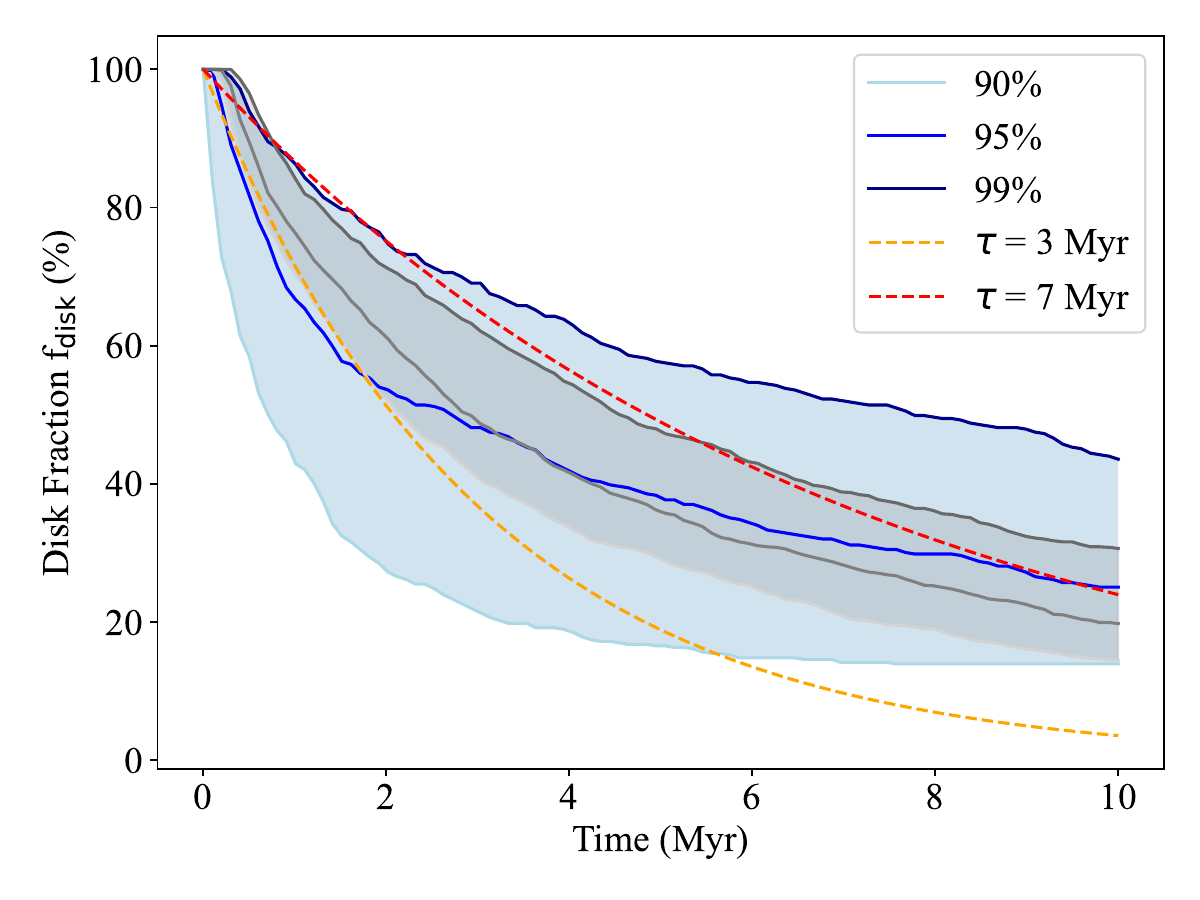} 
  \caption{The decline in disk fraction $f_{\mathrm{disk}}$ with the disk age under different constraints on initial viscosity $\alpha$ and characteristic radius $R_{\mathrm{c}}$.
  The blue and gray only consider disks with $10^{-5} \leq \alpha \leq 10^{-2}$ and $150 \leq R_{\mathrm{c}} \leq 200$\,AU, $10^{-4} \leq \alpha \leq 10^{-2}$ and $10 \leq R_{\mathrm{c}} \leq 30$\,AU, respectively. Others following the same format as Figure \ref{Fig:disk_fraction}.}
  \label{Fig:disk_fraction_150_200}
\end{figure}

\begin{figure}[htbp]
  \centering
  \includegraphics[width=0.47\textwidth]{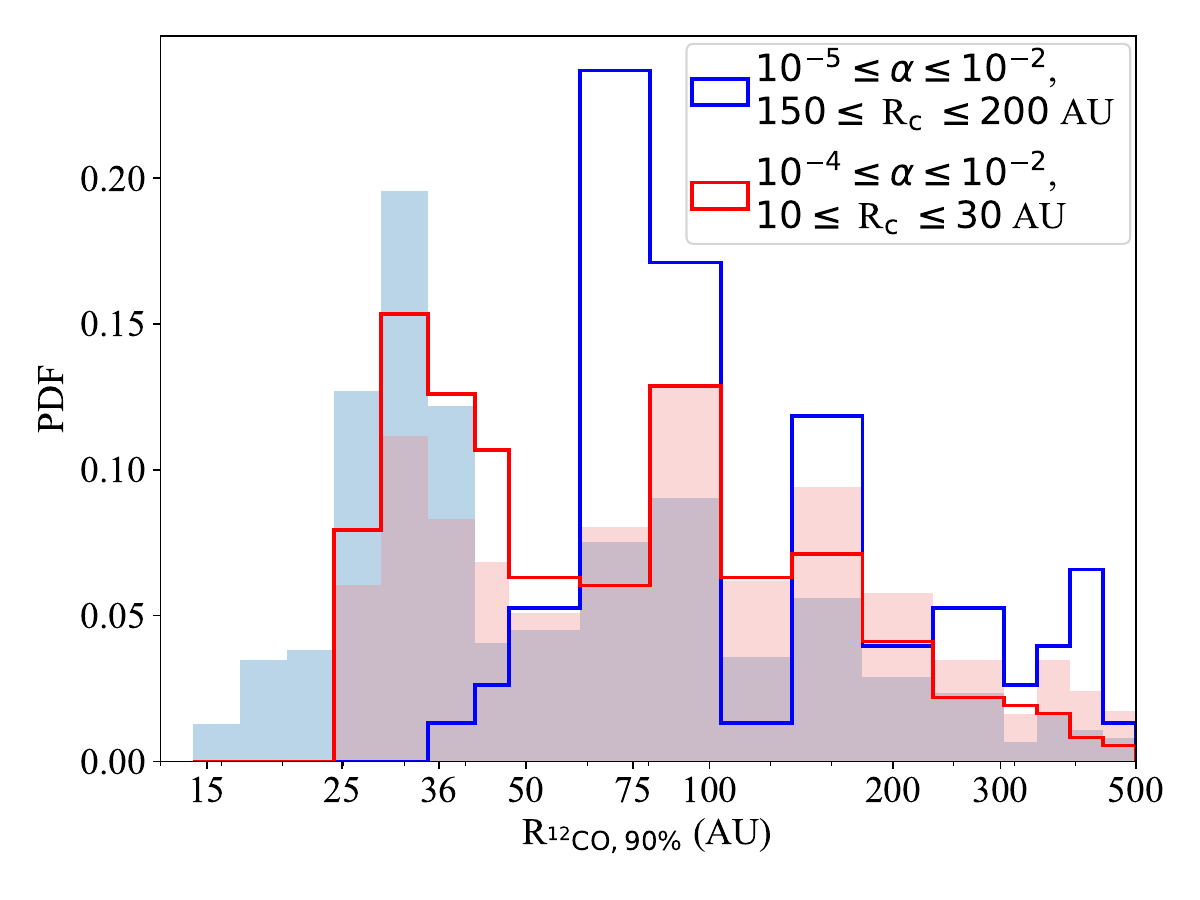} 
  \caption{The PDF of the gas disk radius $R_{^{12}\mathrm{CO}, \, 90\%}$ for surviving disks at 5\,Myr with constraints on initial disk properties. The blue and red represent disks with viscosity $10^{-5} \leq \alpha \leq 10^{-2}$ and $10^{-4} \leq \alpha \leq 10^{-2}$, respectively. The shaded region marks the disks with characteristic radius $10 \leq R_{\mathrm c} \leq 200$\,AU, and solid lines highlight those with $150 \leq R_{\mathrm c} \leq 200$\,AU (blue) and $10 \leq R_{\mathrm c} \leq 30$\,AU (red).}
  \label{Fig:rco_comparison_try}
\end{figure}

\end{appendix}

\end{document}